\documentclass[aps, superscriptaddress,amsmath,amssymb,floatfix, notitlepage, bm,braket, prb, twocolumn, longbibliography]
{revtex4-1}
\pdfoutput=1
\usepackage{times}
\usepackage{graphics}
\usepackage{graphicx}
\usepackage{color}
\usepackage{bm}
\usepackage{braket}
\usepackage{mathtools}
\usepackage{mathcomp}
\usepackage[caption=false,justification=raggedright,position=top,singlelinecheck=off]{subfig}
\usepackage{soul}
\usepackage{tabularx}
\usepackage{calrsfs}
\usepackage[vcentermath]{youngtab}
\usepackage[mathscr]{euscript}
\usepackage[cp1252]{inputenc}
\usepackage[normalem]{ulem}

\newcommand{\enni}{\noindent}
\newcommand{\enbe}{\begin{equation}}
\newcommand{\enee}{\end{equation}}
\newcommand{\enba}{\begin{align}}
\newcommand{\enea}{\end{align}}

\begin{document}

\title{Theory of $(s+id)$ pairing in mixed-valent correlated metals
}
\author{Emilian M.\ Nica}
\affiliation{Department of Physics,
Arizona State University, 
Box 871504, Tempe 85287-1504, AZ, USA}
\email[Corresponding author: ]{enica@asu.edu}
\author{Onur Erten}
\affiliation{Department of Physics,
Arizona State University, 
Box 871504, Tempe 85287-1504, AZ, USA}
\date{\today}

\begin{abstract}
Motivated by the recent discovery of superconductivity in square-planar nickelates as well as by longstanding puzzling experiments in heavy-fermion superconductors, we study Cooper pairing between correlated $d$-electrons coupled to a band of weakly-correlated electrons. We perform self-consistent large N calculations on an effective $t-J$ model for the $d$-electrons with additional hybridization. Unlike previous studies of mixed-valent systems, we focus on parameter regimes where both hybridized bands are relevant to determining the pairing symmetry. For sufficiently strong hybridization, we find a robust $s+id$ pairing which breaks time-reversal and point-group symmetries in the mixed-valent regime. Our results illustrate how intrinsically multi-band systems such as heavy-fermions can support a number of highly non-trivial pairing states. They also provide a putative microscopic realization of previous phenomenological proposals of $s+id$ pairing and suggest a potential resolution to puzzling experiments in heavy-fermion superconductors such as U$_{1-x}$Th$_x$Be$_{13}$ which exhibit two superconducting phase transitions and a full gap at lower temperatures.
\end{abstract}

\maketitle

\section{Introduction}
\label{Sec:Intr}

The historic search for superconductivity (SC) in Ni-based oxides~\cite{Middey_ARMR2016} recently passed an important milestone with Sr-doped NdNiO$_2$~\cite{Hwang_Nature2019, Sawatzky_Nature2019, Zheng_2020}. This long-term pursuit in the nickelates has been partly driven by the similarities with the cuprates. Indeed, both nickelates and cuprates are quasi two-dimensional and share a  nominally half-filled $d_{x^2-y^2}$ orbital with $3d^9$ configuration. However, unlike the typical cuprate parent compound, NdNiO$_2$ is a paramagnetic metal\cite{Hwang_Nature2019} instead of an antiferromagnetic insulator, due to the presence of additional Nd $d$-bands that cross the Fermi level. Self-doping effects push the Ni $d_{x^2-y^2}$ band away from half-filling~\cite{deveraux} and thus from the canonical proximity to the Mott insulator typical of the cuprates. A proper treatment of SC in the nickelates must therefore incorporate the strong correlations of the Ni $d_{x^2-y^2}$ bands and the coupling to weakly interacting Nd $d$-bands alike. In many ways, this parallels heavy-fermion intermetallics\cite{Si_Science2010}, where the strongly-interacting $f$ states couple to weakly-correlated $d$-electrons. Among the intermetallics, actinides with $5f$ orbitals particularly resemble the nickelates since they are closer to the mixed-valent regimes than the lanthanides. They also exhibit superconducting transition temperatures ($T_c$'s) which are comparable to that observed in Sr-doped NdNiO$_2$ as in the case of PuCoGa$_5$ where $T_c \sim $~18.5 K\cite{Curro_Nature2005}. 

\enni
\begin{figure}[ht!]
\centering
\includegraphics[width=0.75\columnwidth]{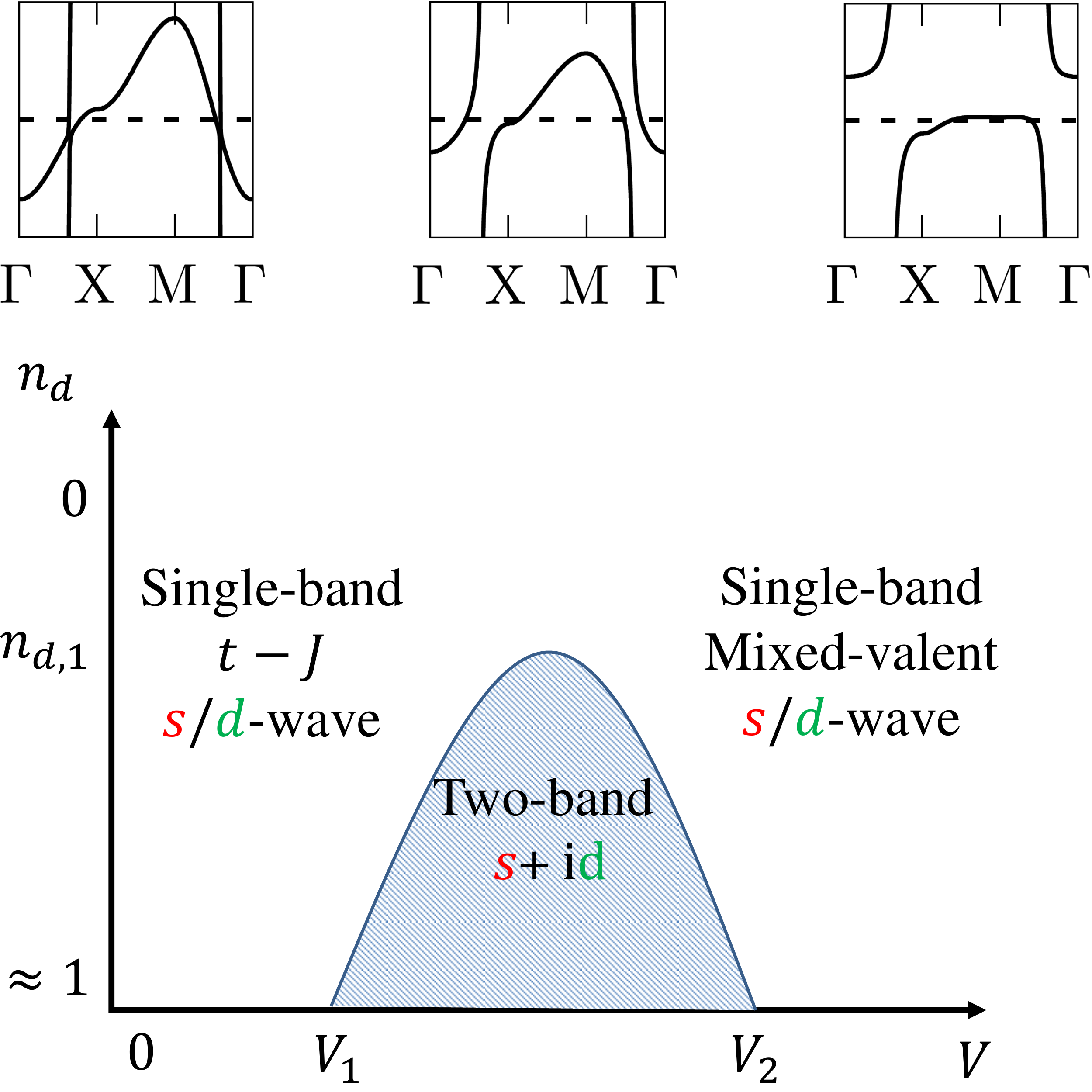}
 \caption{General phase diagram at zero temperature for $s+id$ pairing in a $t-J$ model 
 with a coupling to 
 conduction electrons. $V$ and $n_{d}$ are the hybridization strength and $d$-electron filling, respectively. For simplicity, we consider the case where the hybridization is larger than the $d$-electron hopping $t_{d} \ll V$. When the hybridization is also much weaker than the Heisenberg exchange $V \ll J_{H}$, the pairing instability is determined by a single band with predominant $d$-electron character, as in the $t-J$ model, and the pairing symmetry is typically either $s$- or $d$-wave. In the regime where $V \sqrt{1-n_{d}} \approx J_{H}$, both of the hybridized bands have significant $d$-electron content and $s+id$ emerges as the dominant pairing. Our results are consistent with second-order phase transitions at the boundary of this regime, which is bounded by $V_{1/2}$ and $n_{d,1}$. In the large-$V$ limit, a dominant hybridization gap suppresses pairing on the higher-energy band, resulting in a single-band pairing with either $s$- or $d$-wave symmetry.The upper panels illustrate the general evolution of the hybridized bands in the normal state with increasing $V$. 
 } 
 \label{Fig:Schm}
\end{figure}

To investigate the effects of weakly-correlated conduction $(c)$ electrons on the Cooper pairing of  correlated $d$-electrons, we study a general $t-J_H$ model that includes the usual nearest-neighbor (NN) $d$-electron hopping $t_{d}$ and Heisenberg exchange $J_{H}$, but which also incorporates a $c-d$ hybridization $V$, taken to be local for simplicity.  We perform large N calculations within a Sp(N) representation of the local moments and find that $s+id$ pairing typically occurs in a mixed-valent regime provided that the hybridization is sufficiently strong with $V\sqrt{1-n_{d}} \sim J_{H}$ and $t_{d}/V \ll 1$, where $n_{d}$ is the $d$-electron filling. The emergence of $s+id$ pairing is inherently a two-band phenomenon, requiring a multi-pocketed Fermi surface (FS) with significant $d$-electron content near the Fermi level for both bands. Fig.~\ref{Fig:Schm} summarizes our main conclusions. Our results illustrate the importance of non-trivial multi-band superconductivity, as noted by several proposals for Fe-based~\cite{Yu_2014, Yin_Haule_Kotliar, Coleman, Nica_Yu, Kreisel2017,HY_Hu2019, C_Zhang2013, Sprau} and heavy-fermion~\cite{Pang, Nica_Si, Smidman} SC's. 

$s+id$ pairing is highly unusual as it breaks time-reversal symmetry. Furthermore, it also breaks the point-group symmetry since it is a linear combination of two inequivalent irreducible representations of $D_{4h}$. Consequently, these two channels are not coupled to leading order in a Landau-Ginzburg (LG) theory, and we expect that this type of pairing generally requires two second-order transitions with decreasing temperature. Moreover, $s+id$ pairing vanishes at only four points in the entire Brillouin Zone (BZ) and we expect that the quasiparticle spectrum is typically completely gapped, although accidental point nodes are possible. While our predictions do not strictly apply to the nickelates, where the hybridization between the Ni and the less correlated Nd is typically small, they can provide insight into a number of puzzling experiments in heavy fermion superconductors. One prominent example is the observation of two superconducting phase transitions into a gapped state in U$_{1-x}$Th$_x$Be$_{13}$\cite{Scheidt_PRB1998, White_PhyicaC2015, Stewart_AdvPhys2017, Stewart_2019}. Indeed, Kumar and Wolfle proposed an $s+id$ state for U$_{1-x}$Th$_x$Be$_{13}$ based on a phenomenological LG theory\cite{Kumar_PRL1987}, wheres we here provide a concrete, microscopic realization of such a state. 

In this work, we focus on the mechanisms which can lead to the emergence of $s+id$ pairing in mixed-valent systems rather than an exhaustive study of the phase diagram. In spite of the simplicity of our model, we believe that our main conclusions can also be applied to models with more realistic band-structures. 

Our paper is organized as follows. We discuss the general mechanism behind $s+id$ pairing in Sec.~\ref{Sec:Expl}. In Sec.~\ref{Sec:Mdls}, we introduce our model within an the Sp(N) representation of the spins, derive the mean-field self-consistency equations and BdG spectrum, and discuss the U(1) gauge symmetry. Our numerical results for several regimes are presented in Sec.~\ref{Sec:Rslts}. We discuss our main conclusions in Sec.~\ref{Sec:Dscs}.

\section{$s+id$ pairing in a two-band regime}
\label{Sec:Expl}

Our model for physical correlated $(d)$ electrons is mapped onto an effective $f$-fermion model with renormalized parameters within a slave-boson mean-field theoretical approach. The pairing of these $f$-fermions is determined self-consistently via a standard gap equation, in conjunction with remaining mean-field parameters. Hence, the notions of FS and density of states (DOS) defined for the auxiliary $f$ electrons are crucial in understanding why this type of pairing occurs, although they need not have an immediate physical meaning. Similarly, $f-f$ pairing is not a U(1) gauge-invariant quantity and is thus not a physical observable. The Bogoliubov-de Gennes (BdG) spectrum determined from our effective model is nonetheless physically meaningful. $s+id$ pairing for the $f$ fermions typically leads to a gapped BdG spectrum while breaking time-reversal symmetry, features which remain relevant for the model defined in terms of the physical $d$-electron.  

In order to understand how $s+id$ emerges in our effective model for the $f$ fermions, we first consider the $t-J$ limit with $V=0$. We define an effective kinetic energy scale which is determined for arbitrary doping by the renormalized hopping $t_{d} b^{2}$ and by the contribution of the particle-hole (p-h) channel introduced via the nearest-neighbor (NN) Heisenberg exchange terms. $b=\sqrt{1-n_{f}}$ is the amplitude of the condensed boson in terms of the $d$-electron filling $n_{d}=n_{f}$. We can likewise define a scale associated with the pairing interactions $\sim J_{H}$. In the \emph{absence} of the effective kinetic terms at half-filling, where $b=0$ and the contribution of the p-h terms can be gauged away, the $f$ band becomes flat, leading to $s_{x^{2}+y^{2}}$ and $d_{x^{2}-y^{2}}$ pairing channels with identical critical temperature $T_{c}$~\cite{Kotliar_s_id}, defined via the gap equation and not representing physical superconductivity. Furthermore, $s+id$ pairing emerges at zero-temperature for $t_{d}/J_{H} \lessapprox 1$ within a finite range of hole-doping, provided that the p-h contribution to the kinetic energy is ignored~\cite{Sachdev_Read}. Whenever these p-h terms are taken into account, the quasi-degeneracy between $s$- and $d$-waves is lifted, and $s+id$ pairing is consequently suppressed~\cite{Kotliar}. 

In our model, the hybridization between the $d$- and $c$-electrons introduces an additional scale. In the canonical mixed-valent limit where $t_{d}, J_{H} \ll V$, the scale associated with the hybridization is determined essentially by the indirect hybridization gap $\sim (Vb)^{2}$~\cite{Riseborough_2016}, as depicted in the right top panel in Fig.~\ref{Fig:Schm}. Since the indirect gap exceeds both the effective kinetic and pairing $(\sim J_{H})$ scales, $f-f$ pairing is determined by the single FS due to the lower band~\cite{Coleman_Andrei} and the degeneracy between $s$- or $d$-waves is typically lifted, leading to a suppression of $s+id$ pairing. When $(Vb)^{2}$ determines the lowest scale, we recover the $t-J$ model limit discussed previously. We find that in an intermediate regime where $(Vb)^{2}$ is comparable to both effective kinetic energy and pairing strength scales or, equivalently, when the FS and $f$ fermion content of both bands are relevant to the pairing instability, $s+id$ becomes the preferred state at zero temperature. 

$s+id$ emerges in these regime via two effects apparent in the normal state for $T>T_{c}$. First, due to the hybridization, both $c$-like and $f$-like bands support $f-f$ pairing, in contrast to the single-band limits of the $t-J$ and simple mixed-valent cases, respectively. As the hybridization increases, the FS of the $f$-like band expands while that of the $c$-like band contracts. The $f$-like FS remains in the vicinity of the nodes of the $s_{x^{2}-y^{2}}$ form factor and thus promotes $d_{x^{2}-y^{2}}$ pairing instead. The $c$-like FS, which is centered on the $\Gamma$ point, promotes $s_{x^{2}-y^{2}}$ since the $d_{x^{2}-y^{2}}$ form factor vanishes at $(0,0)$. These two sectors of the FS thus promote two distinct channels, leading a coexisting $s+id$ state. The FS shape here is strongly reminiscent of the Fe-pnictides, where FS pockets present at the edges and center of the BZ~\cite{Lee_Zhang_Wu} can lead to degenerate $s_{\pm}$ and $d_{x^{2}-y^{2}}$ channels. A similar mechanism  persists even when the pocket at $\Gamma$ vanishes with increasing hybridization, provided that the indirect gap remains comparable to the strength of the pairing interactions. Similar effects were noted in early works on superconductivity in Kondo-lattice systems~\cite{Coleman_Andrei, Andrei_Coleman} as well as in the Fe-based SC's~\cite{Goswami}.

Secondly, we observe that the $f$ fermion DOS becomes increasingly concentrated near the Fermi level in the regime where $s+id$ pairing dominates. This is partially a natural consequence of increasing hybridization, as the $c$-like band which overlaps with the $f$-like band in energy, gets pushed to higher energies. However, we also observe an additional effect, as the self-consistent p-h mean-field parameter $K_{\bm{\hat{x},\hat{y}}}$ is also suppressed with increasing hybridization. This leads to a reduction of the effective kinetic energy, and thus to a flattening of the bands which promotes degenerate $s$- and $d$-waves in a way analogous to the $t-J$ model with p-h channel suppressed~\cite{Sachdev_Read}. 

The arguments for $s+id$ pairing have been based  on the properties of the normal state near the pairing $f-f$ critical temperature $T_{c}$. Below $T_{c}$, the pairing, p-h contributions as well as all of the remaining self-consistent parameters can change reflecting the self-consist nature of the calculation~\cite{Kotliar}. However, we find that whenever the normal state reflects the phenomenology discussed thus far, $s+id$ pairing remains the preferred state at zero temperature. 

We characterize various regimes via six independent dimensionless parameters, three of which are defined in terms of the bare coupling constants $\left(t_{d}, V, t_{c}\right)/J_{H}$, where $t_{c}$ is the $c$-electron NN hopping. The remaining parameters reflect temperature $T/J_{H}$, and the $d$-electron and total fillings $n_{d}, n_{Tot}$, respectively. For simplicity, we fix $J_{H}, t_{c}$, and $T$ thereby reducing the number of independent parameters to four. $s+id$ generally emerges in the $n_{d} \lessapprox 1$ regime where the kinetic energy controlled by $\sim t_{d}(1-n_{d})/J_{H}$ is sufficiently suppressed to allow significant $f$ DOS near the Fermi level. In addition, we find that $  V(1-n_{d})/J_{H}$ should approach a value close to unity in order for both bands to be located near the Fermi level. In this regime we also find that smaller values of $t_{d}/V$ are favorable to $s+id$, since a larger kinetic energy tends to push the normal state away from the ideal conditions discussed previously. As we illustrate below, while $s+id$ requires some amount of tuning, it does emerge for a  finite range of dimensionless parameters near $d$ half-filling.    

\section{Models and Solution Methods}     
\label{Sec:Mdls}

In this section, we present our model and solution method. Mixed-valent systems are typically challenging to model faithfully due to complex, multi-band structures with different degrees of correlation~\cite{Stewart_AdvPhys2017}. Similarly, the mechanisms behind Copper pairing are generally not well-established~\cite{Stewart_AdvPhys2017}. Consequently, we consider simplified models which capture the salient features of some of these materials without obscuring the mechanisms behind $s+id$ pairing. Our assumptions include (i) a simple square lattice with tetragonal symmetry; (ii) the presence of a single flavor of weakly-correlated conduction $c$-electrons with NN hopping which hybridize to the correlated $d$-electrons together with NN hopping for the latter; (iii) $c$ and $d$ electrons which belong to identical representations of the $D_{4h}$ point group; and (iv) the inclusion of significant NN hopping for the $d$ electrons.     

In view of points (ii) and (iii), we consider a \emph{local} $c-d$ hybridization. We define an appropriate $t-J$ model which includes the effects of the strong local Coulomb repulsion via the exclusion of double-occupancy for $d$-electrons, NN Heisenberg exchange interactions $J_{H}$, and $c-d$ hybridization $V$. Such models interpolate between single-band $t-J$ and canonical mixed-valent systems, where the dispersion of the correlated $d$ electrons is typically ignored. 

We generalize the SU(2) symmetry of the local spin operators to Sp(N) symmetry in order to obtain a controlled saddle-point solution in the limit of large $N$~\cite{Flint_Dzero_Coleman, Sachdev_Read}. We introduce a slave-boson representation, decouple the exchange interaction in both p-h and p-p channels, and solve these models at saddle-point level \emph{at fixed total filling}. Consequently, the $d$-electron filling is not fixed a priori, but is determined self-consistently. We are interested in solutions which preserve the translational symmetry of the lattice and therefore ignore cases exhibiting phase separation~\cite{Vojta_Sachdev}. 

\subsection{Model}

We consider the following $t-J$ model with a local $c-d$ hybridization:

\begin{widetext}

\enni \begin{align}
H= & P_{d} \bigg[ -2t_{d} \sum_{\braket{ij}, \sigma} \left(d^{\dag}_{i \sigma} d_{j \sigma } +h.c. \right) + \left( \epsilon_{d}- \mu \right)
 \sum_{i, \sigma} n_{di\sigma} 
- t_{c} \sum_{\braket{ij}, \sigma} \left( c^{\dag}_{i\sigma} c_{j \sigma} + h.c. \right) 
+ \left( \epsilon_{c} - \mu \right) \sum_{i , \sigma} n_{ci \sigma}  
 \notag \\
 + & \sqrt{2} V  \sum_{i,\sigma} \left( d^{\dag}_{i \sigma} c_{i \sigma} + h.c. \right) + \frac{J_{H}}{2} \sum_{\braket{ij}} \left( \bm{S}_{di} \cdot \bm{S}_{dj} -\frac{1}{4} n_{di} n_{dj} \right) 
\bigg] P_{d}.
\label{Eq:Lcl_Hmlt}
\end{align}

\end{widetext}

\enni $P_{d}$ is the projection of doubly-occupied $d$-electron states, the $i,j$ indices cover the square lattice, $\sigma$ represents the spins of the electrons, and $\epsilon_{c/d}$ are on-site energies. Similarly, 

\enni \begin{align}
\bm{S}_{di} = & \frac{1}{2} \sum_{\alpha \beta}  d^{\dag}_{i\alpha} \bm{\sigma}_{\alpha \beta} d_{i \beta},
\\ 
n_{di\sigma} = & d^{\dag}_{i\sigma}d_{i\sigma},
\\
n_{di}= & \sum_{\sigma} n_{di\sigma},
\end{align}

\enni are the SU(2) spin, spin-resolved, and total $d$-electron filling operators respectively. Analogous  definitions hold for the $c$-electrons. The $d$-electron hopping $t_{d}$, $c-d$ hybridization $V$, and NN Heisenberg interactions $J_{H}$ have been re-scaled for convenience. 

The model can be formally derived by projecting out double-occupancy within a Hubbard model for the $d$-electrons, while allowing the additional hybridization with the $c$-electrons. Although such a procedure also generates additional Kondo exchange interactions, these are much smaller than the hybridization $V$, and shall therefore be ignored.  

In the limit $V \rightarrow 0$, the Hamiltonian in Eq.~\ref{Eq:Lcl_Hmlt} reduces to a single-band $t-J$ model. In the limit $t_{d} \rightarrow 0$, the Hamiltonian reduces to a standard mixed-valent model with additional exchange interactions and $d$-electron hopping. For general parameters, $H$ interpolates between these two limits.

Following Refs.~\onlinecite{Flint_Dzero_Coleman, Sachdev_Read}, we generalize the Hamiltonian from SU(2) to symplectic Sp(N) symmetry by promoting all of the spin indices $\sigma$ to Sp(N) indices $p$ and by subsequently replacing the Heisenberg exchange interaction with the corresponding expression using Sp(N) generators 

\enni \begin{align}
& \frac{J_{H}}{2N} \sum_{\braket{ij}} \sum_{pq} \sum_{\alpha \beta} \sum_{\gamma \delta} S^{pq}_{\alpha \beta} S^{qp}_{\gamma \delta} d^{\dag}_{i \alpha} d_{i \beta} d^{\dag}_{j \gamma} d_{j \delta} 
\notag \\
= & -  \frac{J_{H}}{N} \sum_{pq} \bigg\{ 
d^{\dag}_{i p} d_{j p} d^{\dag}_{j q} d_{i q} + \tilde{p}\tilde{q} d^{\dag}_{i \bar{q}}  d^{\dag}_{j q} d_{j p} d_{i \bar{p}}\bigg\} 
\notag \\
- & \frac{J_{H}}{N} \sum_{p} d^{\dag}_{i p} d_{i p}.
\label{Eq:SpN_exch}
\end{align}

\enni The Sp(N) generators are~\cite{Flint_Dzero_Coleman} 

\enni \begin{align}
S^{pq}_{\alpha \beta} = & \delta^{p}_{\alpha} \delta^{q}_{\beta} - \tilde{\alpha} \tilde{\beta} \delta^{p}_{-\beta} \delta^{q}_{-\alpha},
\end{align}

\enni where the indices $p,q \in [\pm 1, \pm N ]$ with $N$ even and

\enni \begin{align}
\tilde{p} = \text{sgn}(p).
\end{align}

\enni We re-scaled the Heisenberg exchange in order to obtain a finite contribution in the large $N$ limit. For similar reasons, we rescale $t_{d} \rightarrow t_{d}/N$ and $V \rightarrow V/\sqrt{N}$. 

\subsection{Saddle-point solutions in the large-$N$ limit }

Following Ref.~\onlinecite{Vojta_Sachdev} we introduce a slave-boson representation for the $d$ electrons 

\enni \begin{align}
d_{i p} \rightarrow b^{\dag}_{i} f_{ip}
\end{align}

\enni together with the local constraint

\enni \begin{align}
b^{\dag}_{i} b_{i} + \sum_{p}f^{\dag}_{i p} f_{i p} = \frac{N}{2}, 
\label{Eq:Gtzw}
\end{align}

\enni enforced via a Lagrange multiplier $\lambda$. 

In addition, we impose a fixed total filling 

\enni \begin{align}
\sum_{i, p} \left( f^{\dag}_{ip} f_{ip} + c^{\dag}_{ip} c_{ip} -  \frac{n_{Tot}}{2} \right) =0,
\label{Eq:ntt}
\end{align}

\enni where $0 \le n_{Tot} \le 4$. This condition is enforced via a chemical potential $\mu$. Recall that, unlike the standard single-band $t-J$ model~\cite{Kotliar, Sachdev_Read}, here the $d$-electron filling $n_{di}=n_{fi}$ is not fixed but is determined self-consistently. 

We ignore the density-density interactions present in the original model at SU(2) symmetry~\cite{Vojta_Sachdev}. Furthermore, we decouple the NN exchange interactions in Eq.~\ref{Eq:SpN_exch} in both p-h and particle-particle (p-p) channels via the \emph{dimensionless} parameters

\enni \begin{align}
K_{i, \bm{e}} = &  \frac{1}{N} \sum_{q} \braket{ f^{\dag}_{\bm{r}_{i} + \bm{e}, q} f_{\bm{r}_{i}, q} } 
\label{Eq:K}
\\
B_{i, \bm{e}} = & \frac{1}{N} \sum_{q} \tilde{q} \braket{f_{\bm{r}_{i}, \bar{p}} f_{\bm{r}_{i}+\bm{e}, p}},
\label{Eq:B}
\end{align}

\enni where $\bm{e} \in \{\bm{\hat{x}}a , \bm{\hat{y}}a \}$ are the NN lattice vectors.

We consider solutions at saddle-point level which preserve the translation symmetry of the lattice. As our model has a U(1) gauge symmetry~\cite{Vojta_Sachdev}, we choose a gauge where the condensed boson is real and uniform:

\enni \begin{align}
\braket{b^{\dag}_{i}} = \braket{b_{i}} = \left( \sqrt{\frac{N}{2}} \right) b.
\end{align}

\enni where $b$ is also independent of the lattice site, as are the p-h and p-p dimensionless parameters defined previously.

The LG action per site is 

\begin{widetext}

\enni \begin{align} 
f = &  \sum_{p > 0} \bigg\{ \frac{1}{N_{s}} \sum_{\bm{k}}  \left[ -2T \sum_{m}  \ln\left( \cosh(\beta E_{\bm{k}m})  \right) + \text{Tr}\left[h_{\bm{k}} \right] + 2J_{H}\sum_{\bm{e}} \left[  B^{\dag}_{\bm{e}} B_{\bm{e}} + \chi_{\bm{e}} \chi^{\dag}_{\bm{e}} \right] 
+  \lambda  \left( b^{2} - 1 \right)+  \mu n_{Tot} \right] \bigg\},
\end{align}

\end{widetext}

\enni where $\beta=1/T$ and $E_{\bm{k}m}$ are the eigenvalues of 

\enni \begin{align}
H_{\bm{k}} = \begin{pmatrix}
\hat{h}_{\bm{k}}  & \hat{\Delta}_{\bm{k}} \\
\hat{\Delta}^{\dag}_{\bm{k}}  & - \hat{h}^{T}_{-\bm{k}} 
\end{pmatrix}
\label{Eq:BdG_Hmlt} 
\end{align}

\enni in a Nambu basis with spinor 
$\Psi^{T}=(c_{\bm{k p}}, f_{\bm{k} p}, c^{\dag}_{-\bm{k} \bar{p}} f^{\dag}_{-\bm{k} \bar{p}} )$. The normal part is given by 

\enni \begin{align}
\hat{h}_{\bm{k}} = 
\begin{pmatrix}
\epsilon_{\bm{k}c} & Vb \\
Vb & \epsilon_{\bm{k}f}
\end{pmatrix},
\label{Eq:h_ht}
\end{align}

\enni where 

\enni \begin{align}
\epsilon_{\bm{k}c} = & -2t_{c} \sum_{\bm{k}} \cos\left(\bm{k \cdot e} \right) + \left[ \epsilon_{c} - \mu \right] 
\\
\epsilon_{\bm{k}f} = &  -2t_{d} b^{2}    
\sum_{\bm{k}} \sum_{\bm{e}} \cos\left( \bm{k \cdot e} \right) 
\\ 
- & 2J_{H} \left[  K^{'}_{\bm{e}} \cos( \bm{k \cdot e}) - K^{''}_{\bm{e}} \sin(\bm{k \cdot e} ) \right] + \left[ \epsilon_{d} - \mu + \lambda \right].
\end{align}

\enni As the solutions presented here involve $K^{''}_{\bm{e}}=0$, we define an effective $f$-fermion hopping as

\enni \begin{align}
t_{f, \bm{e}} = t_{d}b^{2} + J_{H} K^{'}_{\bm{e}}.
\label{Eq:Effc_kntc}
\end{align}

\enni $t_{f,\bm{e}}$ defines an \emph{effective kinetic energy scale} as discussed in Sec.~\ref{Sec:Expl}. 

The pairing part of $H_{\bm{k}}$ is determined by 

\enni \begin{align}
\hat{\Delta}_{\bm{k}} = &
\begin{pmatrix}
0 & 0 \\
0 & \Delta_{\bm{k}}
\end{pmatrix},
\label{Eq:Dlt_ht}
\end{align}

\enni where 

\enni \begin{align}
\Delta_{\bm{k}} = & - 2 J_{H }\sum_{\bm{e}} B_{\bm{e}}  \cos( \bm{k \cdot e}).
\end{align}

\enni We also define the complex dimensionless pairing mean-field parameters corresponding to the $s_{x^{2}+y^{2}}$ and $d_{x^{2}-y^{2}}$ channels as

\enni \begin{align}
B_{s} = & B_{x} + B_{y} 
\notag \\
= & \left| B_{s} \right| e^{i\phi_{s}} 
\label{Eq:Bs}
\\
B_{d} = & B_{x} - B_{y} 
\notag \\
= & \left| B_{d} \right| e^{i\phi_{d}}.
\label{Eq:Bd}
\end{align}

\enni The relative phase is defined as 

\enni \begin{align}
\phi_{Relative} = \phi_{s}-\phi_{d}.
\end{align}

Saddle-point solutions are obtained in standard fashion via variation of the self-consistent parameters $\lambda, K_{\bm{e}}, B_{\bm{e}}, \mu$, and $b$. The self-consistency conditions are given in Eqs.~\ref{Eq:Gtzw},~\ref{Eq:ntt},~\ref{Eq:K},~\ref{Eq:B} together with  

\enni \begin{align}
\left[ -4t_{d}  \sum_{\bm{e}} \text{Re} K_{\bm{e}} + \lambda \right] b + \frac{2V}{N}\sum_{p} \text{Re} \braket{f^{\dag}_{\bm{r_{i}}p} c_{\bm{r_{i}} p}} =0. 
\end{align}

\enni In the $t_{d}=0$ limit, this equation reduces to the standard mixed-valent case~\cite{Coleman_1987}. In the $V=0$ limit, we recognize the condition for boson condensation in a typical $t-J$ model~\cite{Kotliar}. 

We obtain self-consistent solutions numerically on a $100 \times 100$ square lattice. In practice, we tune the $d$-electron level $\epsilon_{d}$ with all other parameters fixed such that both $b$ and $n_{f}$ are determined self-consistently. 

\subsection{Pairing and gauge invariance}

The action corresponding to the Hamiltonian in Eq.~\ref{Eq:Lcl_Hmlt} in the Sp(N) slave-boson the formulation is invariant under a U(1) gauge transformation~\cite{Read_Newns, PA_Lee_2006}:

\enni \begin{align}
b_{i} \rightarrow & b_{i} e^{i \theta_{i}} \\
f_{ip} \rightarrow & f_{ip} e^{i \theta_{i}} \\
\lambda_{i} \rightarrow & \lambda_{i} - i \dot{\theta}_{i} \\
K_{i,j} = & K_{i,j} e^{i(\theta_{j}-\theta_{i})} \\
B_{i,j} = & B_{i,j} e^{i(\theta_{i}+\theta_{j})}.
\end{align}

\enni The self-consistent dimensionless $f-f$ Hartree (p-h) $K_{i,\bm{e}}$ and paring $B_{\bm{e}}$ (p-p) parameters defined in Eq.~\ref{Eq:B} are not invariant under arbitrary U(1) gauge transformations. However, the associated quasiparticle spectrum is real and observable, and thus is invariant under the U(1) gauge transformation defined above. 

\subsection{Quasiparticle spectrum for $s+id$ pairing}

The spectrum of the BdG Hamiltonian in Eq.~\ref{Eq:BdG_Hmlt}, which determines the physical quasiparticle spectrum is given by 

\begin{widetext}
\begin{align}
E_{\bm{k}\pm} = \pm \sqrt{\frac{\epsilon^{2}_{\bm{k}c}
+ \epsilon^{2}_{\bm{k}f} + 2(Vb)^{2} + \Delta^{2}_{\bm{k}s} + \Delta^{2}_{\bm{k}d} \pm \sqrt{\left[ (\epsilon^{2}_{\bm{k}f} - \epsilon^{2}_{\bm{k}c}) + \Delta^{2}_{\bm{k}s} + \Delta^{2}_{\bm{k}d}\right]^{2} + 4(Vb)^{2} \left[(\epsilon_{\bm{k}f} + \epsilon_{\bm{k}c})^{2} +(\Delta_{\bm{k}s} + \Delta_{\bm{k}d})^{2} \right]}
}{2}},
\label{Eq:BdG_dspr}
\end{align}

\end{widetext}

\enni where we used

\enni \begin{align}
\Delta_{\bm{k}} = \Delta_{\bm{k}s} + i \Delta_{\bm{k}d},
\end{align}

\enni with $\Delta_{\bm{k}s/d}$ real. Note the inner square-root which is due to the non-commuting  matrices $\hat{h}_{\bm{k}}$ and $\hat{\Delta}_{\bm{k}}$ defined in Eqs.~\ref{Eq:h_ht} and~\ref{Eq:Dlt_ht}. Consequently, the gap is not $|\Delta_{\bm{k}}|^{2}$ as in a simple one-band case~\cite{Nica_Yu}. The spectrum reverts to an effective single-band $t-J$ case when $(Vb)^{2} \ll \Delta^{2}_{\bm{k}s} + \Delta^{2}_{\bm{k}d}$ with a conventional gap on the $f$ band. Similarly, in the opposite limit $(Vb)^{2} \gg \Delta^{2}_{\bm{k}s} + \Delta^{2}_{\bm{k}d}$, we recover the spectrum of weak-coupling pairing occurring predominantly on the lower-energy $f$-like band. 

As both $\Delta_{\bm{k}s/d}$ vanish at $(\pm \pi/2, \pm \pi/2)$, nodes at $E_{\bm{k}\pm}=0$ can appear provided that these points are on the FS. More generally, due to the two-band nature of $s+id$ pairing, which is reflected in the unconventional form of the gap, it is possible that nodes can emerge away from $\pi/2, \pi/2$ even though $\Delta^{2}_{\bm{k}s} + \Delta^{2}_{\bm{k}d}$ remains finite. This is due to the additional inner-square root term in Eq.~\ref{Eq:BdG_dspr}, which can compensate the remaining factors.

\section{Results}
\label{Sec:Rslts}

We use a convention whereby all coupling constants appearing in our model have arbitrary units of energy. Without loss of generality, we set the $c$-electron NN TB coefficient $t_{c}$ and on-site energies $\epsilon_{c}$ to 0.5 and 0, respectively.   

We first illustrate the emergence of $s+id$ pairing from the $t-J$ limit under increasing the local $c-d$ hybridization $V$. We present our results for two values of the fixed $d$-electron hopping $t_{d}=0.01$ and $0.1$ at fixed total filling $n_{Tot}=1.473$, NN Heisenberg exchange $J_{H}=0.0375$, and $c$-electron NN hopping $t_{c}=0.5$, while varying $V$ for a finite range of $d$-electron filling $n_{d}$. We subsequently consider cases with smaller $n_{Tot}=1.16$. Our results illustrate that $s+id$ pairing occurs for a set of different parameters and hence that this unconventional pairing state does not require fine-tuning.

As the filling of the physical $d$ electrons coincides with that of the auxiliary $f$ fermions within the slave-boson approximation i.e. $n_{d}=n_{f}$, we shall use the two naming conventions interchangeably. All of our results correspond to self-consistent solutions with condensed boson $b \neq 0$.

\subsection{$t_{d}=0.01, J_{H}=0.0375$, and $n_{Tot}=1.473$}
\label{Sec:IIA}

\begin{figure}[t!]
\centering
\includegraphics[width=1.0\columnwidth]{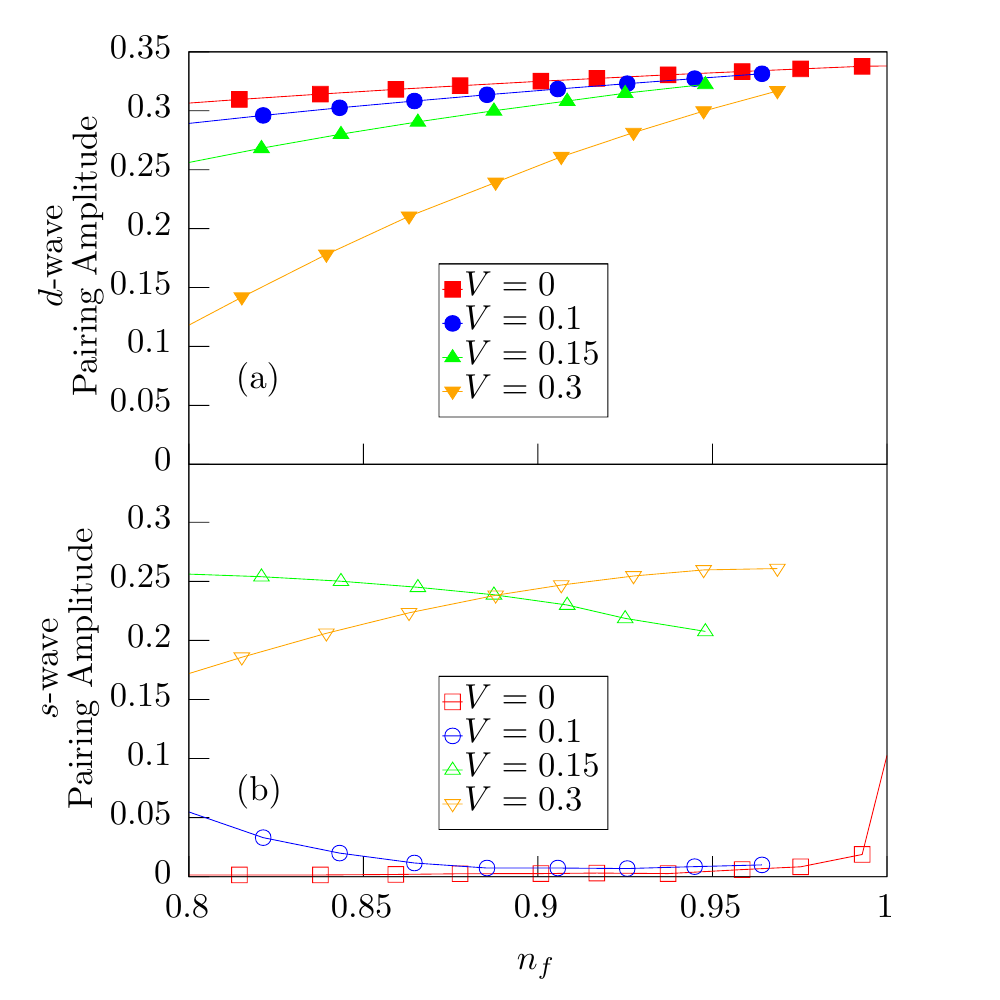}
 \caption{Amplitudes of the dimensionless $f-f$ pairing amplitudes (Eqs.~\ref{Eq:Bd} and ~\ref{Eq:Bs}) for the $d_{x^{2}-y^{2}}$ and $s_{x^{2}+y^{2}}$ channels in panels (a) and (b), respectively, as functions of the hybridization $V$ and $d$-electron filling $n_{d}=n_{f}$ for fixed $t_{d}=0.01, J_{H}=0.0375$ and $n_{Tot}=1.473$ at zero temperature. The $s$-wave amplitude is suppressed in the $t-J$ $(V=0)$ limit and becomes finite for $V \ge 0.1$.} 
 \label{Fig:1}
\end{figure}

We consider the case for $t_{d} = 0.01, J_{H}=0.0375$ with fixed $n_{Tot}=1.473$. In this limit, the hopping of the $d$ electrons (see Eq.~\ref{Eq:Effc_kntc}), plays a sub-leading role when compared to the contribution of the p-h mean-field parameter $K_{\bm{e}}$. In Fig.~\ref{Fig:1}, we plot the amplitudes of the $d_{x^{2}-y^{2}}$ and $s_{x^{2}+y^{2}}$, $f-f$ dimensionless pairing mean-field parameters (see Eqs.~\ref{Eq:Bs},~\ref{Eq:Bd}), at zero temperature as functions of the hybridization $V$ and filling $n_{f}$.   

\enni

\enni
\begin{figure}[t!]
\centering
\includegraphics[width=1.0\columnwidth]{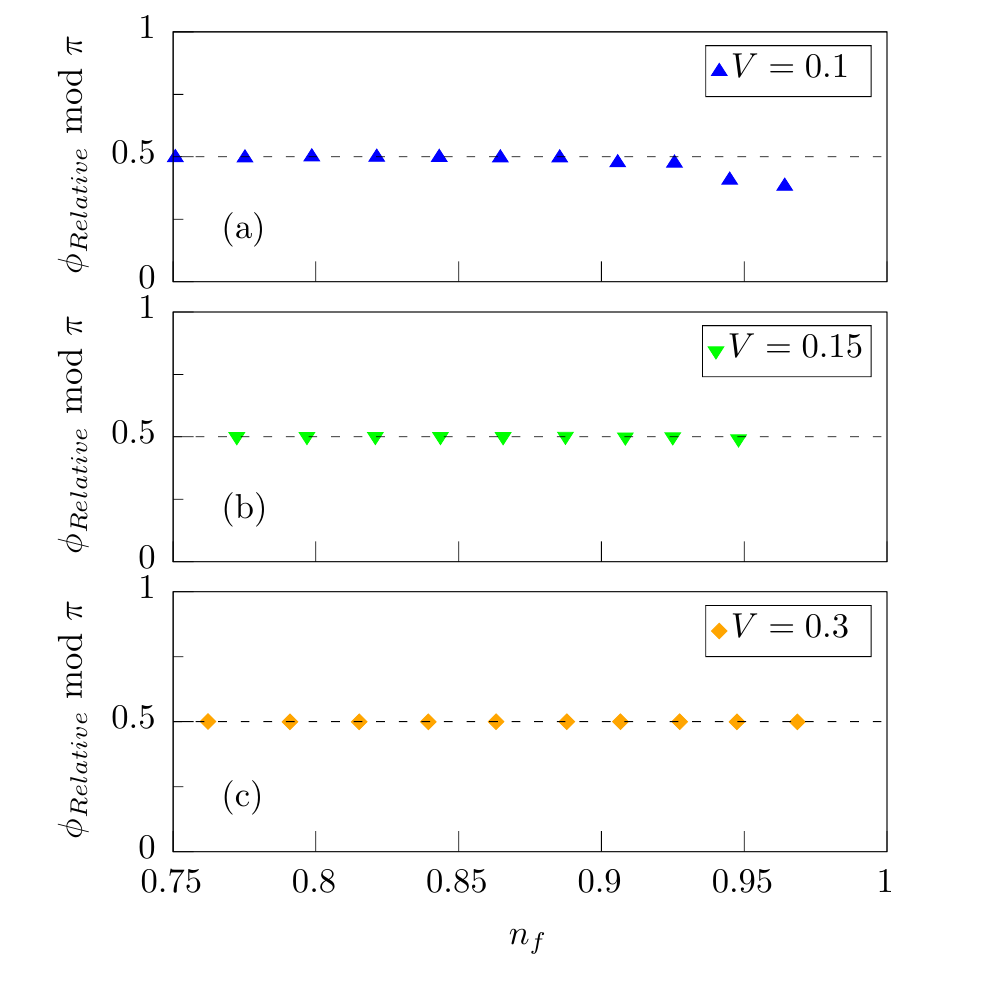}
 \caption{Relative phase $\phi_{Relative}$ (Eqs.~\ref{Eq:Bs} and~\ref{Eq:Bd}) mudulo $\pi$ of $s_{x^{2}+y^{2}}$ and $d_{x^{2}-y^{2}}$ channels as functions of the hybridization $V$ and $n_{f}$ for fixed $t_{d}=0.01, J_{H}=0.0375$ and $n_{Tot}=1.473$, at zero temperature in units of $\pi$. A $\phi_{Relative}=\pi/2$ indicates that $s$- and $d$-wave pairing coexist in an $s+id$ pairing state.}
 \label{Fig:2}
\end{figure}

\enni As shown in panel (a), the $d$-wave amplitude remains finite throughout the entire range of $n_{f}$, although it does suffer a reduction due to the increasing hybridization which pushes the system away from the strong-coupling limit of the $t-J$ model. By contrast, the $s$-wave amplitude is almost completely suppressed in the $t-J$ limit $(V=0)$ (red symbols) but becomes finite for $V \ge 0.1$. To illustrate that the $s$- and $d$-wave components are locked into a $s+id$ pairing state at $T=0$, in Fig.~\ref{Fig:2} we plot the relative phase $\phi_{Relative}$ (Eqs.~\ref{Eq:Bs} and~\ref{Eq:Bd}) of the $s$- and $d$-wave channels as functions of $V$ and $n_{f}$, modulo $\pi$ in units of $\pi$ for a subset of the parameters of Fig.~\ref{Fig:1}. It is apparent that a $\pi/2$ relative phase persists whenever both $s$- and $d$-wave amplitudes are finite and therefore, that $s+id$ pairing indeed emerges at zero temperature. The $s+id$ state persists for larger values of $V$, albeit within a reduced range of $n_{f} \approx 1$. 

\begin{figure}[t!]
\centering
\includegraphics[width=1.0\columnwidth]{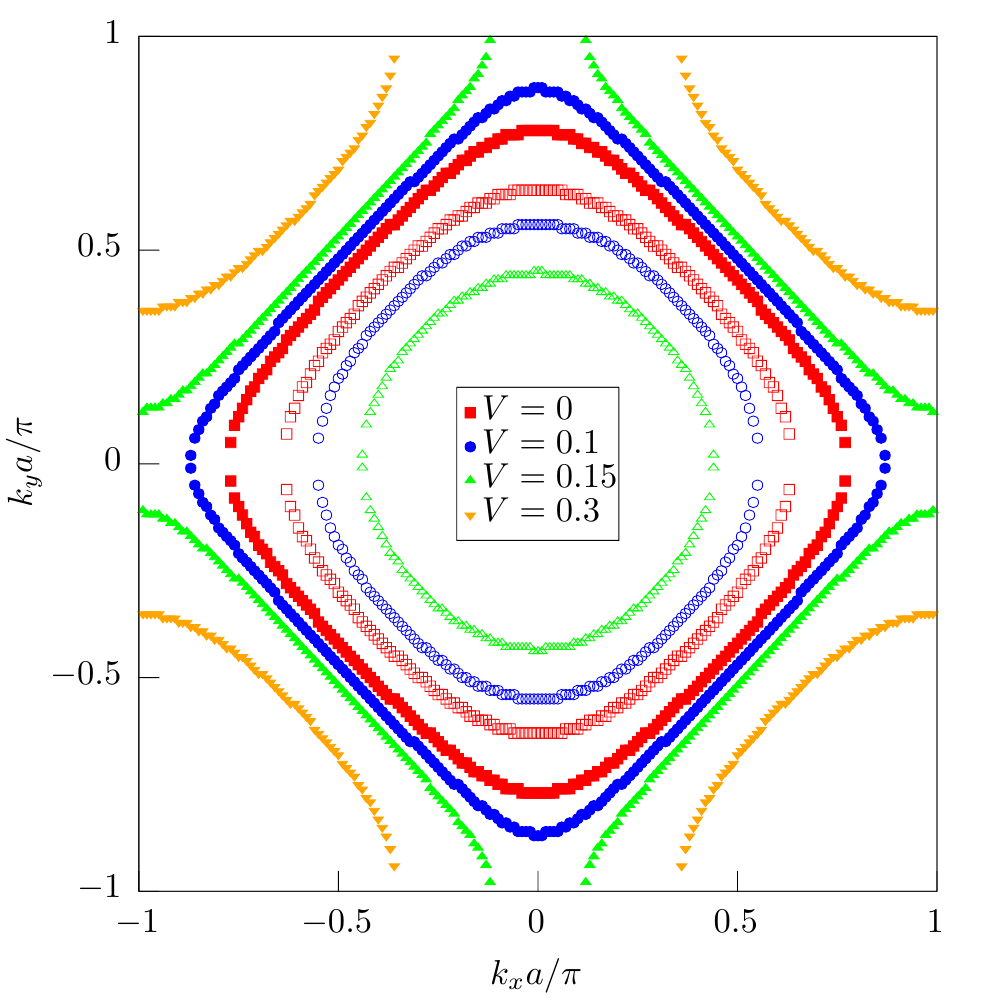}
 \caption{FS's determined at $T=0.001$, where the pairing is strongly suppressed, as functions of the hybridization $V$ at fixed $t_{d}=0.01, J_{H}=0.0375, n_{f}=0.9$, and $n_{Tot}=1.473$, covering the same parameter regime of Fig.~\ref{Fig:1}. The red symbols represent the $t-J~(V=0)$ limit, with the $f$ fermion  and $c$-electron FS's indicated by the solid and hollow squares, respectively. Increasing $V$ tends to enlarge one of the FS sectors (solid symbols) while shrinking the other (hollow symbols). This change in the FS likewise tends to promote $s$- and $d$-wave pairing, as described in the text. Note the absence of the pocket at $\Gamma$ for $V=0.3$. Both bands still support $f-f$ pairing, and therefore promote $s+id$ pairing, as the indirect gap is smaller than the pairing interactions $\sim J_{H}$.}
 \label{Fig:3}
\end{figure}

As discussed previously, each of the distinct FS sectors which emerge under increasing hybridization from the $t-J$ limit promote $s$- and $d$-waves, respectively. To illustrate this mechanism, we consider the normal state at finite temperature $T=0.001$ determined from self-consistent solutions with vanishing pairing amplitudes and $b \neq 0$. In Fig.~\ref{Fig:3}, we plot the evolution of the FS's as functions of $V$ for fixed $n_{f}=0.9$. At $V=0$, the larger FS (red solid squares) corresponds to the purely $f$-fermion band while the smaller pocket (red hollow squares) is comprised entirely of $c$-electrons. As $V$ is increased, the larger pockets (solid symbols) grow and move beyond the "diamond" shape typical at $n_{f}=1$, while the smaller pockets (hollow symbols) shrink and eventually vanish. The $s$-wave pairing amplitude (Fig.~\ref{Fig:1}~(b)) becomes comparable to the $d$-wave amplitude once the FS moves away from the $(0, \pi)$ to $(\pi,0)$ lines as shown by the green and orange symbols in Fig.~\ref{Fig:3}. We also note that both $V=0.15$ and $V=0.3$ cases exhibit comparable $s$- and $d$-wave amplitudes although the Fermi pocket at $\Gamma$ is absent in the latter case. As mentioned previously, this is due to the presence of an indirect gap which is smaller than the pairing interactions $\sim J_{H}$ ensuring that both bands still support $f-f$ pairing and that they  promote $s$- and $d$-wave as in the two-sector FS cases. 

\begin{figure}[t!]
\centering
\includegraphics[width=1.0\columnwidth]{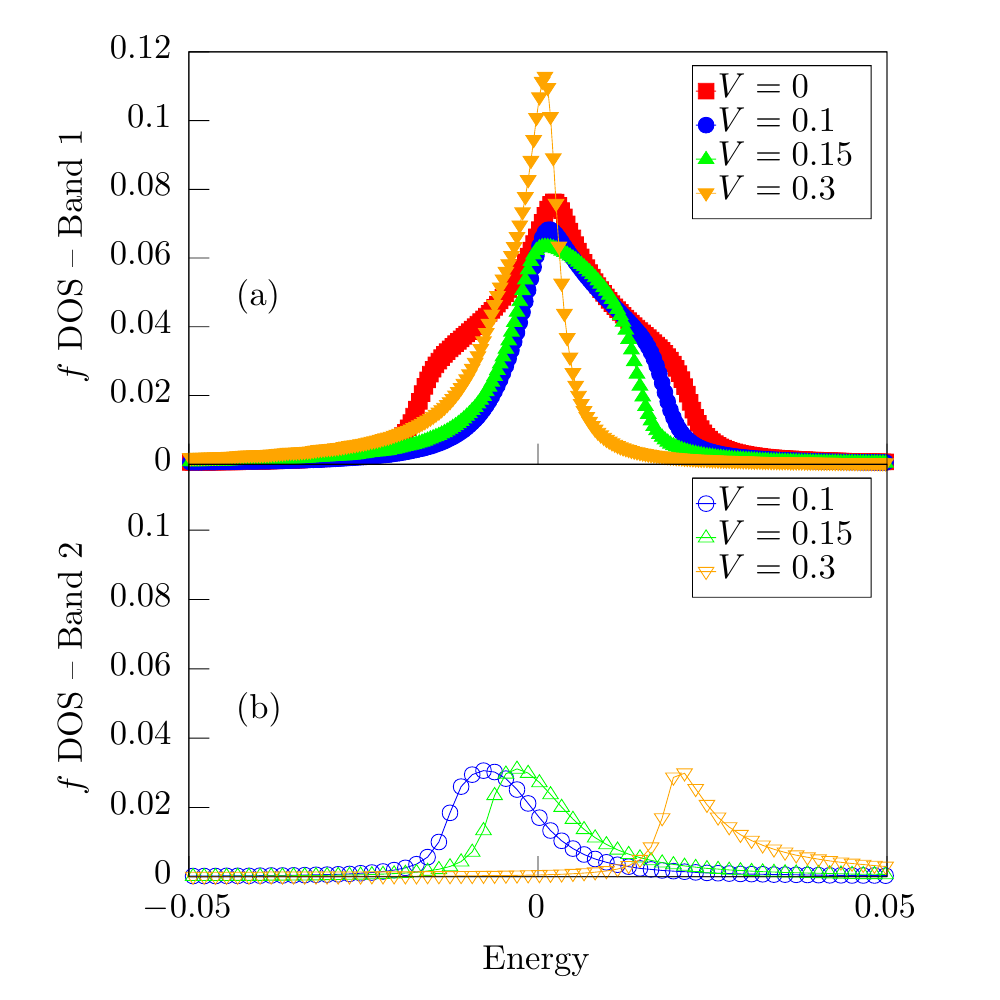}
 \caption{$f$ DOS in the normal state at $T=0.001$ projected onto each of the two bands as functions of energy and $V$ at fixed $t_{d}=0.01, J_{H}=0.0375, n_{f}=0.9$, and $n_{Tot}=1.473$. A uniform broadening is applied in all cases. The chemical potential is pinned to zero energy. The DOS is illustrated for a single band in the $t-J$ limit (red symbols). A narrowing of the peaks is apparent with increasing $V$ culminating with the $V=0.3$ case (orange symbols). The dramatic sharpening of the peaks in this case is due to a change in sign of the self-consistent p-h parameter $K$ which strongly renormalizes the kinetic energy due to hopping ($\sim t_{d}$, Eq.~\ref{Eq:Effc_kntc}). The hybridization gap in this case is filled due to the artificial broadening.}
 \label{Fig:4}
\end{figure}

Alongside the FS, the $f$-fermion DOS in the normal state, which reflects the combined contributions of the hopping $(\sim t_{d})$ and p-h terms $(\sim J_{H})$~(Eq.~\ref{Eq:Effc_kntc}) to the effective kinetic energy and the effects of hybridization $(\sim V)$, also plays an important role in the emergence of $s+id$ pairing. A reduction in the widths of the $f$ DOS peaks promotes degenerate $s$- and $d$-wave pairing, as in the $t-J$ model at half-filling. To illustrate, in Fig.~\ref{Fig:4}, we plot the $f$ DOS projected onto each of the two bands, in the normal state at $T=0.001$, as a functions of $V$, for fixed $t_{d}=0.01, J_{H}=0.0375, n_{f}=0.9$ and $n_{Tot}=1.473$, or in the same parameter regime as Fig.~\ref{Fig:3}. Note that zero-energy corresponds to the Fermi level. A fixed broadening was applied in order to smoothen the curves. The red solid squares in panel (a) illustrate the $f$ DOS in the single-band $t-J$ limit. As $J_{H}$ is the dominant coupling with $t_{d}/J_{H} \lessapprox 0.2$ and $V=0$, the width of the peak is determined mainly by the contribution of the p-h mean-field parameter $K_{\bm{e}}$ (Eq.~\ref{Eq:Effc_kntc}) to the effective kinetic energy. Upon increasing the hybridization to $V=0.1$, the single peak splits into two contributions for each of the bands, reflecting the mixing of $f$ and $c$ electrons. Most of the $f$ weight still resides in the higher-energy band (solid blue squares). Beyond $V=0.15$ (green triangles), the weight on the lower-energy band (green hollow triangles) gets shifted closer to the chemical potential at zero energy. Also note the simultaneous change in the FS, as shown in Fig.~\ref{Fig:3} together with the emergence of $s+id$ pairing as shown in Fig.~\ref{Fig:1}~(b). The shift in the $f$ DOS closer to the Fermi level together with the change in FS mark the crossover from effective single-band pairing in the small $V \ll 0.1$ regime to the intermediate two-band pairing picture discussed illustrated in Fig.~\ref{Fig:Schm} and discussed in Sec.~\ref{Sec:Expl}. For even higher $V=0.3$ (orange symbols), we note a dramatic sharpening of the peak for the lower-energy band and a simultaneous opening of a hybridization gap which is slightly obscured by the artificial broadening. The sharp peak can be attributed to the p-h term which \emph{changes sign w.r.t. $t_{d}$} ensuring an even greater reduction in the effective bandwidth (Eq.~\ref{Eq:Effc_kntc}). The reduction in kinetic energy w.r.t the $t-J$ limit, as demonstrated by the narrowing of the $f$ DOS peaks with increasing hybridization promotes degenerate $s$- and $d$-wave pairing. 

\begin{figure}[t!]
\centering
\includegraphics[width=1.0\columnwidth]{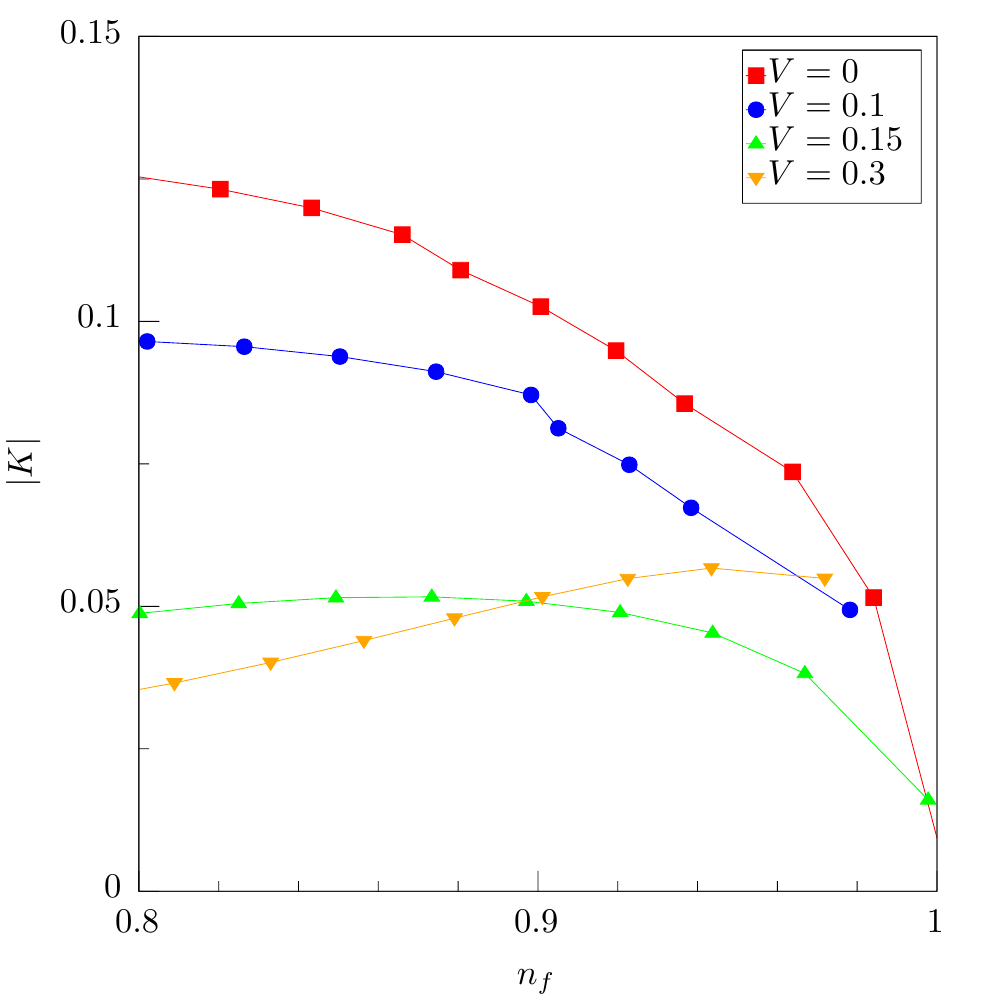}
 \caption{Amplitudes of the p-h mean-field parameter $\left| K \right| = \left| K_{\bm{\hat{x}}} \right| = \left| K_{\bm{\hat{x}}} \right|$ in the normal state at $T=0.001$ as functions of $V$ and $n_{f}$, for fixed $t_{d}=0.01, J_{H}=0.0375$ and $n_{Tot}=1.473$. $\left| K \right|$ decreases with increasing $V$ for $V \le 0.15$. Although $\left| K \right|$ increases from $V=0.15$ to $V=0.3$, the sign of this mean-field parameter also changes within this range, further reducing the effective kinetic energy in Eq.~\ref{Eq:Effc_kntc}.}
 \label{Fig:5}
\end{figure}

To further illustrate the reduction of the effective kinetic energy with increasing hybridization, in Fig.~\ref{Fig:5}, we plot the amplitude of the dimensionless NN p-h mean-field parameter $\left| K \right|=\left| K_{\bm{\hat{x}}} \right|=\left| K_{\bm{\hat{y}}} \right|$ (Eq.~\ref{Eq:K}) in the normal state at $T=0.001$, as a function of $V$ and $f$ filling $n_{f}$ for fixed $t_{d}=0.01, J_{H}=0.05$, and $n_{Tot}=1.473$. $K$ renormalizes the hopping of the $f$ fermions (Eq.~\ref{Eq:Effc_kntc}) and thus controls the effective kinetic energy. It is apparent that $\left| K \right|$ decreases monotonically for $V$ up to 0.15. For $V=0.3$, the amplitude of $K$ actually increases near $n_{f} \approx 1$. However, in contrast to all of the other cases shown here, $K$ changes sign, and is therefore subtracted from rather than added to part proportional to $t_{d}$. This ensures that the $f$ DOS is strongly peaked near the Fermi level, as shown in Fig.~\ref{Fig:4}. 

\enni

\enni

\subsection{$t_{d}=0.1, J_{H}=0.0375$, and $n_{Tot}=1.473$}
\label{Sec:IIB}

In order to illustrate that $s+id$ pairing state can occur in a finite range of $d$-electron filling for larger values of the hopping coefficient $t_{d}$, we present our results for $t_{d}=0.1$, one order of magnitude larger than previously shown in Sec.~\ref{Sec:IIA}, for fixed $J_{H}=0.0375$. 

\begin{figure}[t!]
\centering
\includegraphics[width=1.0\columnwidth]{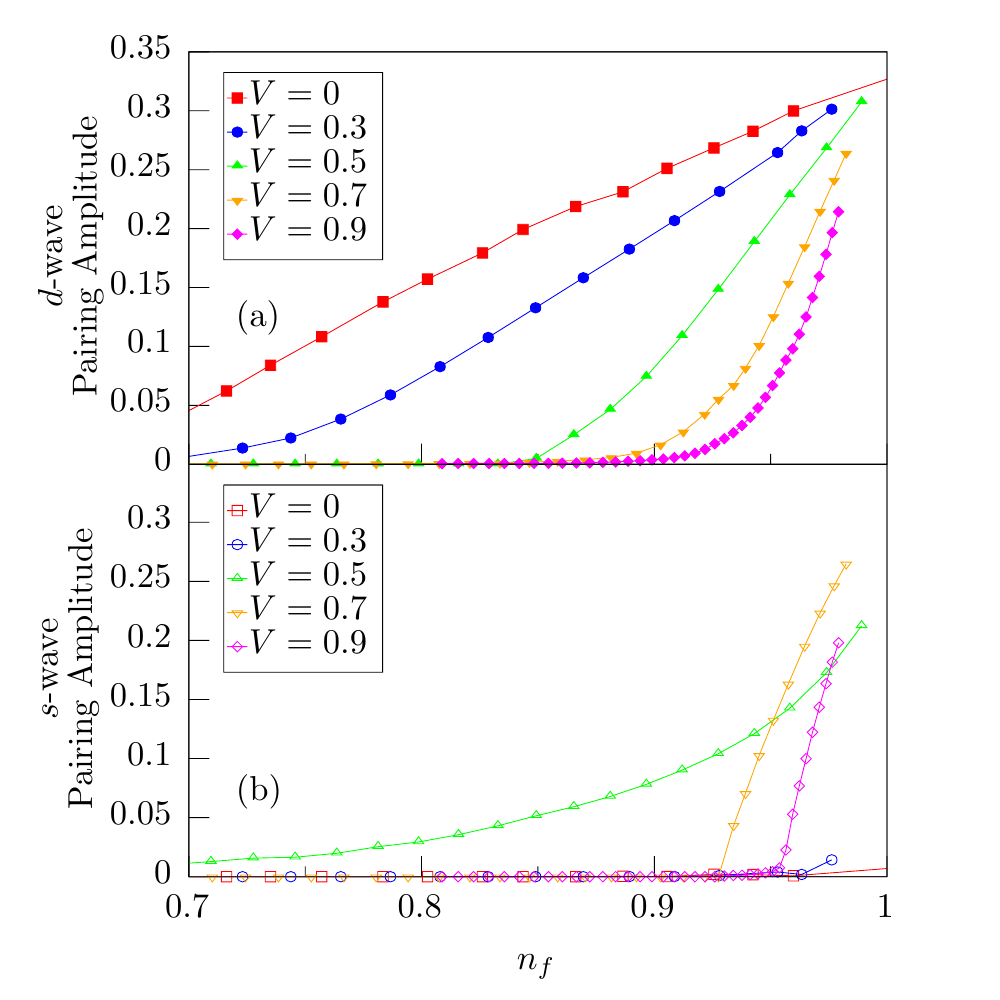}
 \caption{Amplitudes of the dimensionless $f-f$ pairing amplitudes for the $d_{x^{2}-y^{2}}$  and $s_{x^{2}+y^{2}}$ channels in panels (a) and (b), respectively, as functions of the hybridization $V$ and and $n_{f}$ for fixed $t_{d}=0.1, J_{H}=0.0375, n_{Tot}=1.473$ at zero temperature. The $s$-wave amplitude is suppressed in the $t-J$ $(V=0)$ limit and becomes finite for $V \ge 0.5$. Note that for $V=0.5$ and smaller $n_{f}$, $s+id$ pairing is suppressed in favor of a simple $s$-wave.}
\label{Fig:6}
\end{figure}

In Fig.~\ref{Fig:6}, we show the $d$- and $s$-wave pairing amplitudes at $T=0$ as functions of $V$ and $n_{f}$ for fixed total filling $n_{Tot}=1.473$. As for the smaller $t_{d}=0.01$, the $s$-wave amplitude is suppressed in the $t-J$ limit for $V=0$ shown in panel (b) (red, hollow squares). However, it becomes finite for $V \ge 0.5$. The mechanisms behind the emergence of $s+id$ pairing are similar to those for the $t_{d}=0.01$ case of Sec.~\ref{Sec:IIA}, as shown in Appendix~\ref{Sec:Appn_A}. 

Note that for $V=0.5$ and for $nf \lessapprox 0.83$, there is a second-order phase transition from $s+id$ to simple $s$-wave instead of simple $d$-wave, as for the other values of $V$ shown in the Fig.~\ref{Fig:6}. As the $f$ DOS stays relatively constant for a range exceeding the pairing strength $\sim J_{H}$ in the vicinity of the Fermi level, the pairing falls within a weak-coupling regime as shown in Fig.~\ref{Fig:Appn_A_V} in Appendix~\ref{Sec:Appn_A}. The corresponding FS (shown in Fig.~\ref{Fig:Appn_A_IV} of the same Appendix) maintains contributions from both bands even in this limit, in contrast to the cases with $t_{d}=0.01$ discussed in Sec.~\ref{Sec:IIA}. This illustrates that the emergence of $s+id$ pairing requires both a sufficiently strong concentration of $f$ DOS states near the Fermi level and a favorable FS. 

\enni

\subsection{$t_{d}=0.1, J_{H}=0.0375$, and $n_{Tot}=1.16$}
\label{Sec:IIC}

\begin{figure}[t!]
\centering
\includegraphics[width=1.0\columnwidth]{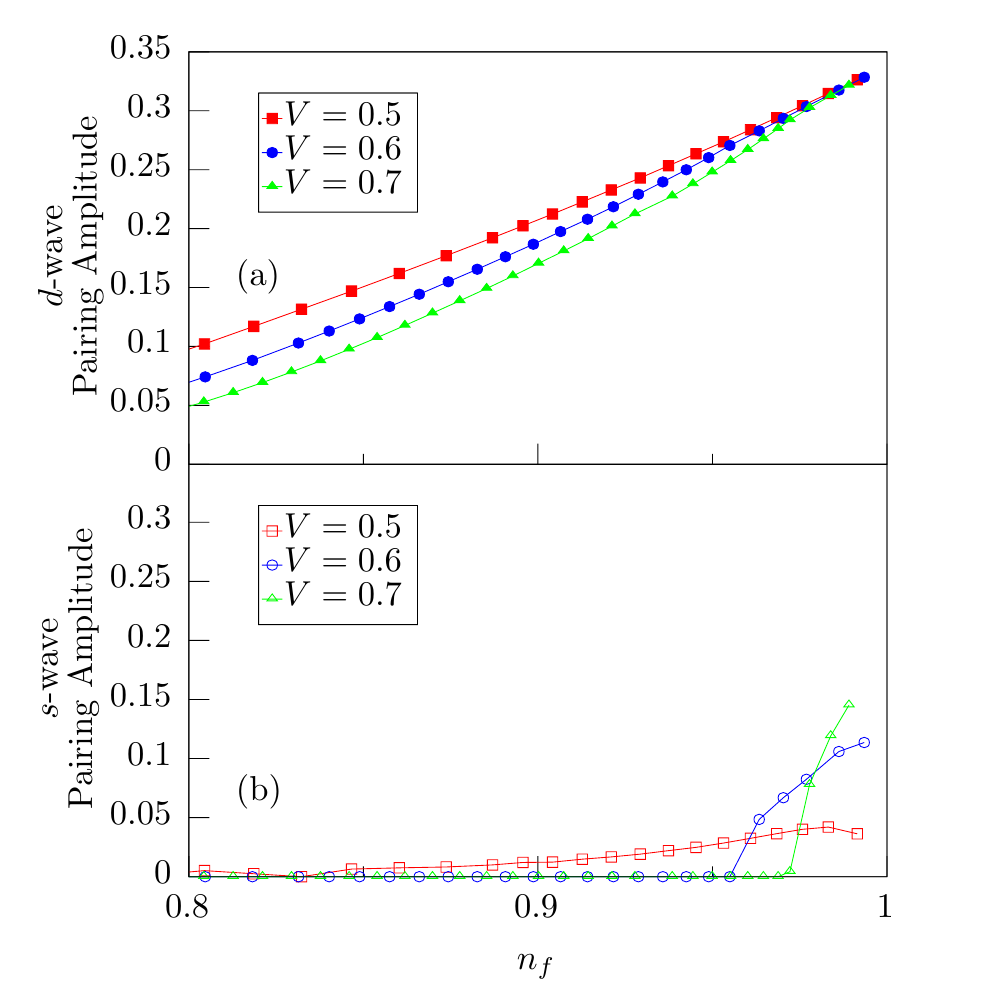}
 \caption{Amplitudes of the dimensionless $f-f$ pairing amplitudes for the $d_{x^{2}-y^{2}}$  and $s_{x^{2}+y^{2}}$ channels in panels (a) and (b), respectively, as functions of the hybridization $V$ and and $n_{f}$ for fixed $t_{d}=0.1, J_{H}=0.0375$ and reduced total filling $n_{Tot}=1.16$ at zero temperature. The $s$-wave amplitude is finite near the $f$ half-filling point but is suppressed when compared to the cases of higher $n_{Tot}=1.473$ previously discussed.}
\label{Fig:7}
\end{figure}

We illustrate that $s+id$ also emerges near the $f$ half-filling point for a range of total fillings $n_{Tot}$. In Fig.~\ref{Fig:7}, we plot the $s$- and $d$-wave amplitudes at $T=0$ for fixed $t_{d}=0.1, J_{H}=0.0375$, and $n_{Tot}=1.16$ as functions of $V$ and $n_{f}$. The $s$-wave amplitude becomes significant near half-filling, although it is suppressed when compared to the results of $n_{Tot}=1.473$ presented in the previous sections. This suppression is a consequence of the reduced size of the FS pocket at the center of the BZ, as the $c$-band is nearer to it's bottom for this regime of smaller total filling. For the same reason, the bare $f$- and $c$-bands are nearer in energy such that the hybridization further depletes the $f$ DOS near the Fermi level. 

\enni

\section{Discussion}
\label{Sec:Dscs}

We studied a simplified $t-J$ model for correlated $d$-electrons which hybridize with weakly-correlated $c$-electrons within a self-consistent mean-field theory with Sp(N) representation of the spins. We found robust $s+id$ pairing extending for a finite range near $d$-electron half-filling, for typical values of $d$-electron hopping and exchange interactions, provided that the hybridization is sufficiently strong. We illustrated that the $s+id$ state in our model is due essentially to a two-band pairing in contrast to previous studies of $t-J$ models where this type of pairing also emerges under doping.  

Our results illustrate that $s+id$ pairing occurs with increasing hybridization from the $t-J$ limit due first to a FS which includes both bands and which favors $s$- and $d$-wave channels to similar extent, and secondly, to an increase in the correlated-electron DOS near the Fermi level. Our results bridge the gap between correlated single $d$-band models, as for the cuprates and mixed-valent systems where pairing occurs mainly in the close vicinity of the FS. As such, our results illustrate how the pairing in correlated multi-band systems can generically exhibit a variety of unconventional phases. 

The rather inclusive conditions leading to the emergence of $s+id$ pairing within our toy model are possible within more realistic treatments of mixed-valent systems such as DFT+DMFT, provided that the multi-band nature of these systems is taken into account. We believe that our results also illustrate how previous phenomenological proposals, as in the case of U$_{1-x}$Th$_x$Be$_{13}$, can be realized microscopically within a generic two-band model. 

During the preparation of this manuscript, we became aware of Ref.~\onlinecite{Zhan_2020} which considers a similar $t-J$ model with additional Kondo interactions within a renormalized mean-field theory in the context of Sr-doped NdNiO$_2$. The authors find an $s+id$ pairing phase in a regime where the Kondo coupling is the highest energy scale and the Kondo-induced hybridization is finite. In this context, we also note the recent single particle tunneling experiments on superconducting nickelate thin films~\cite{Gu_2020} which find spectra consistent with two distinct pairing symmetries, one which is naturally associated with a $d$-wave and another which exhibits a full gap. While tentative at this stage, we find that these studies further hint at the possibility that non-trivial pairings such as $s+id$ are not uncommon in systems with pronounced mixed-valent character. 

\begin{acknowledgements}
We thank Sumilan Banerjee and Qimiao Si for useful discussions related to this work. EN is supported by ASU startup grant and OE is supported by NSF-DMR-1904716. We acknowledge the
ASU Research Computing Center for HPC resources.
\end{acknowledgements}

\appendix

\section{Normal state for $t_{d}=0.1, J_{H}=0.0375$ and $n_{Tot}=1.473$}
\label{Sec:Appn_A}

In this section we present the normal-state properties for the case with fixed $t_{d}=0.1, J_{H}=0.0375$ and $n_{Tot}=1.473$ discussed in Sec.~\ref{Sec:IIB}.

\begin{figure}[h!]
\centering
\includegraphics[width=1.0\columnwidth]{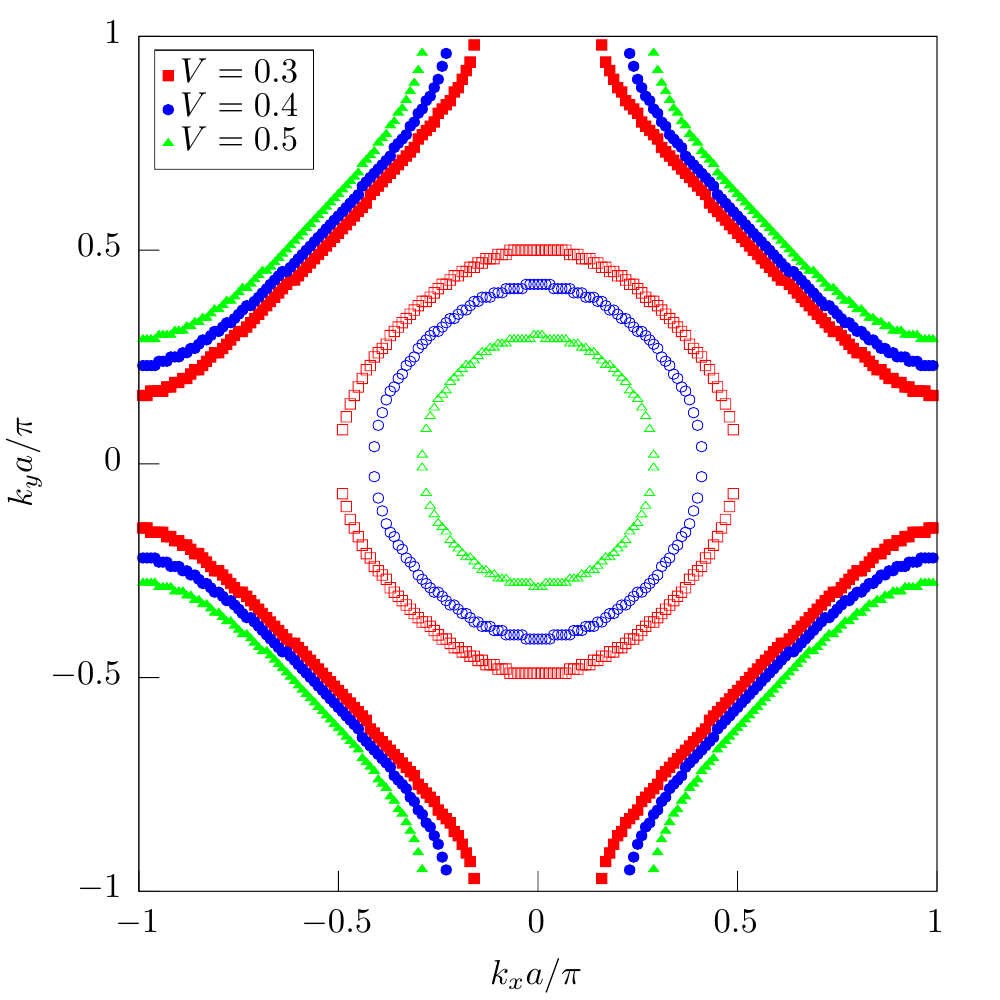}
\caption{FS's determined in the normal state at $T=0.001$ as functions of $V$ for fixed $t_{d}=0.1, J_{H}=0.0375, n_{f}=0.9$, and $n_{Tot}=1.473$. The solid symbols are adiabatically connected to the $d$- and $c$-electron bands in the $V=0$ limit. The FS's evolve with increasing $V$ in a way analogous to the cases with smaller $t_{d}=0.01$ shown in Fig.~\ref{Fig:3}.}
\label{Fig:Appn_A_I}
\end{figure}

In Fig.~\ref{Fig:Appn_A_I}, we present the FS's determined at $T=0.001$ and $n_{f}=0.9$ as functions of $V$. Upon increasing $V$ we observe that the sectors closer to the $M$ points (solid symbols) are growing while those centered on the $\Gamma$ point are shrinking, mirroring the case with $t_{d}=0.01$ shown in Fig.~\ref{Fig:3}. 
\enni

\begin{figure}[h!]
\centering
\includegraphics[width=1.0\columnwidth]{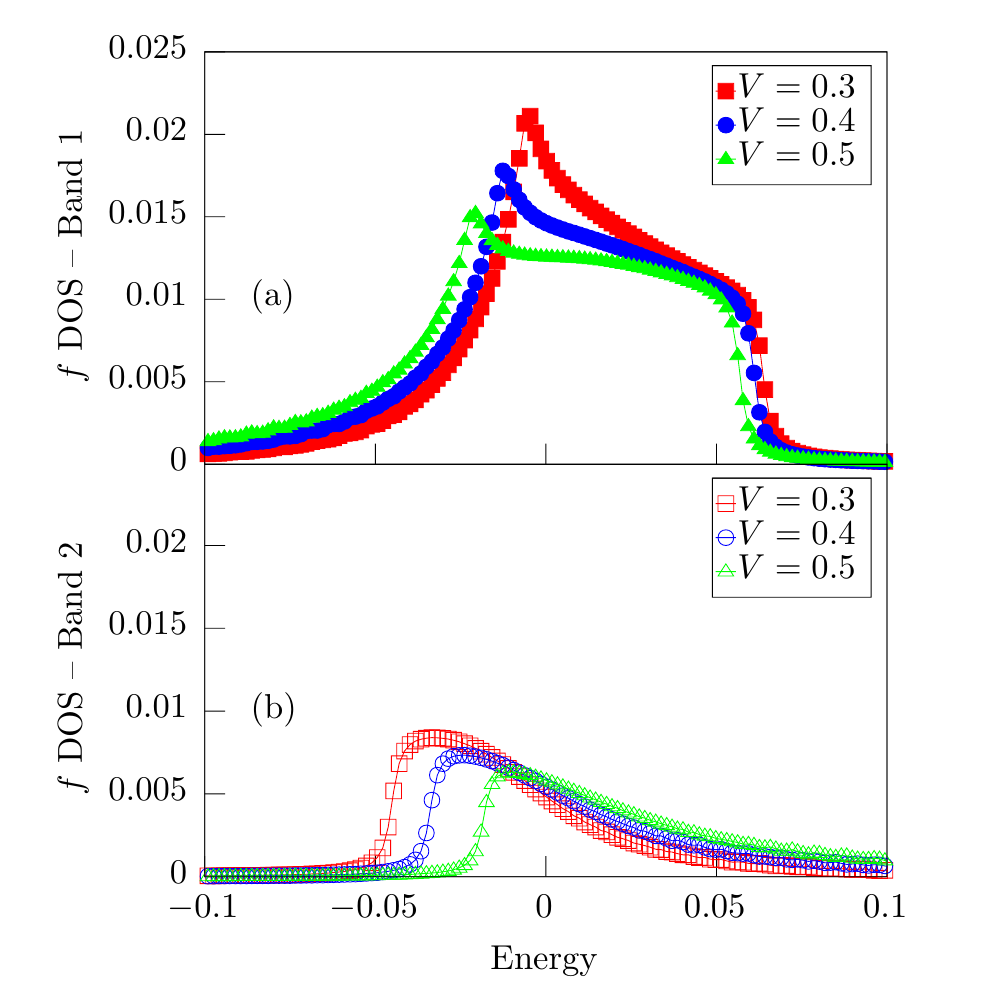}
\caption{$f$ DOS projected onto the two bands determined in the normal state $T=0.001$ as functions of energy and $V$ for fixed $t_{d}=0.1, J_{H}=0.0375, n_{f}=0.9$, and $n_{Tot}=1.473$. The Fermi level is pinned to zero-energy. The narrowing of the peaks observed for $t_{d}=0.01$ in Fig.~\ref{Fig:4} is obscured here by the much larger value of $t_{d}=0.1$.}
\label{Fig:Appn_A_II}
\end{figure}

\begin{figure}[h!]
\centering
\includegraphics[width=1.0\columnwidth]{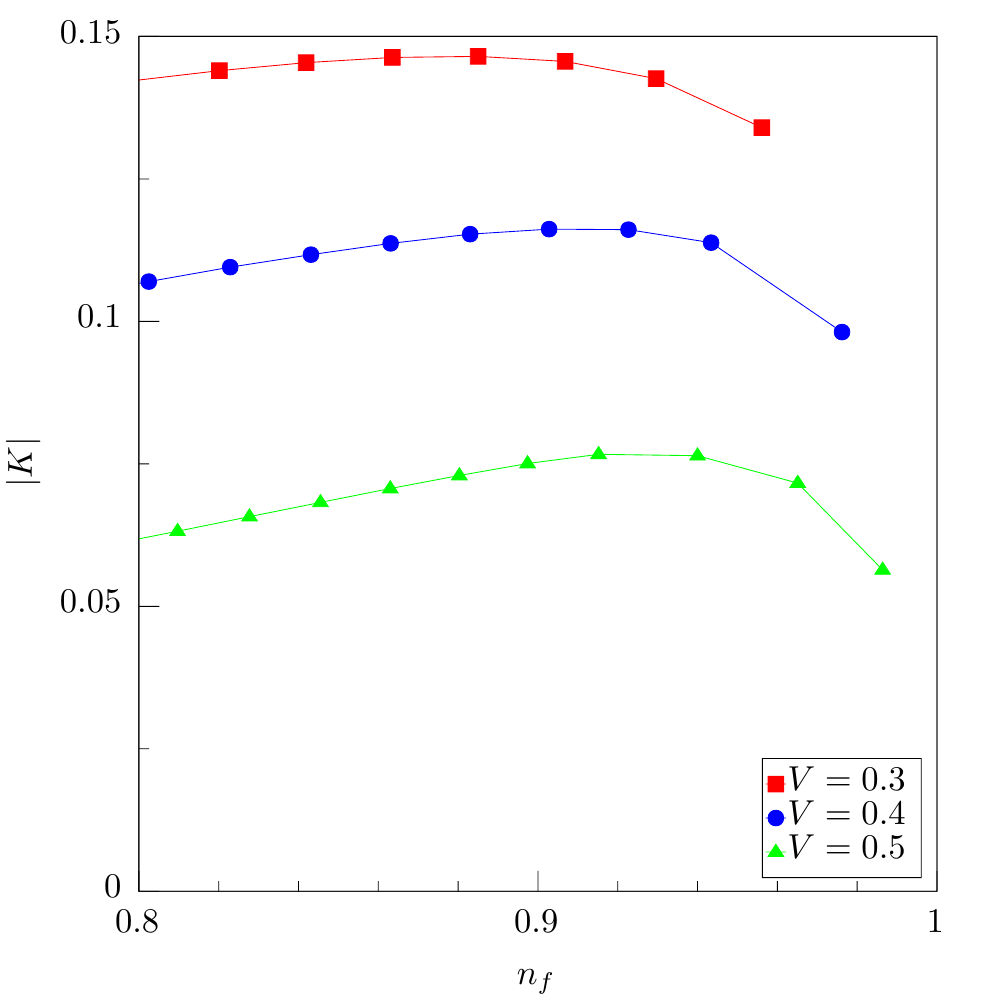}
\caption{Amplitudes of the self-consistent p-h terms in the normal state at $T=0.001$ as functions of $V$ and $n_{f}$ for fixed $t_{d}=0.1, J_{H}=0.0375$, and $n_{Tot}=1.473$. The amplitudes decrease monotonically with $V$.}
\label{Fig:Appn_A_III}
\end{figure}

In Fig.~\ref{Fig:Appn_A_II}, we show the $f$ DOS projected onto the two bands for the same parameter range. We observe a gradual shift in the DOS of band 2 toward the Fermi level at zero-energy with increasing hybridization. Note that the sharpening of the peaks for $t_{d}=0.01$ (Fig.~\ref{Fig:4}) which occurs due to a renormalization of the p-h term is obscured here by the much larger contribution of $t_{d}=0.1$ to the effective kinetic energy scale although the amplitude of the p-h term is suppressed here as well, as shown in Fig.~\ref{Fig:Appn_A_III}.  
\enni

\begin{figure}[h!]
\centering
\includegraphics[width=1.0\columnwidth]{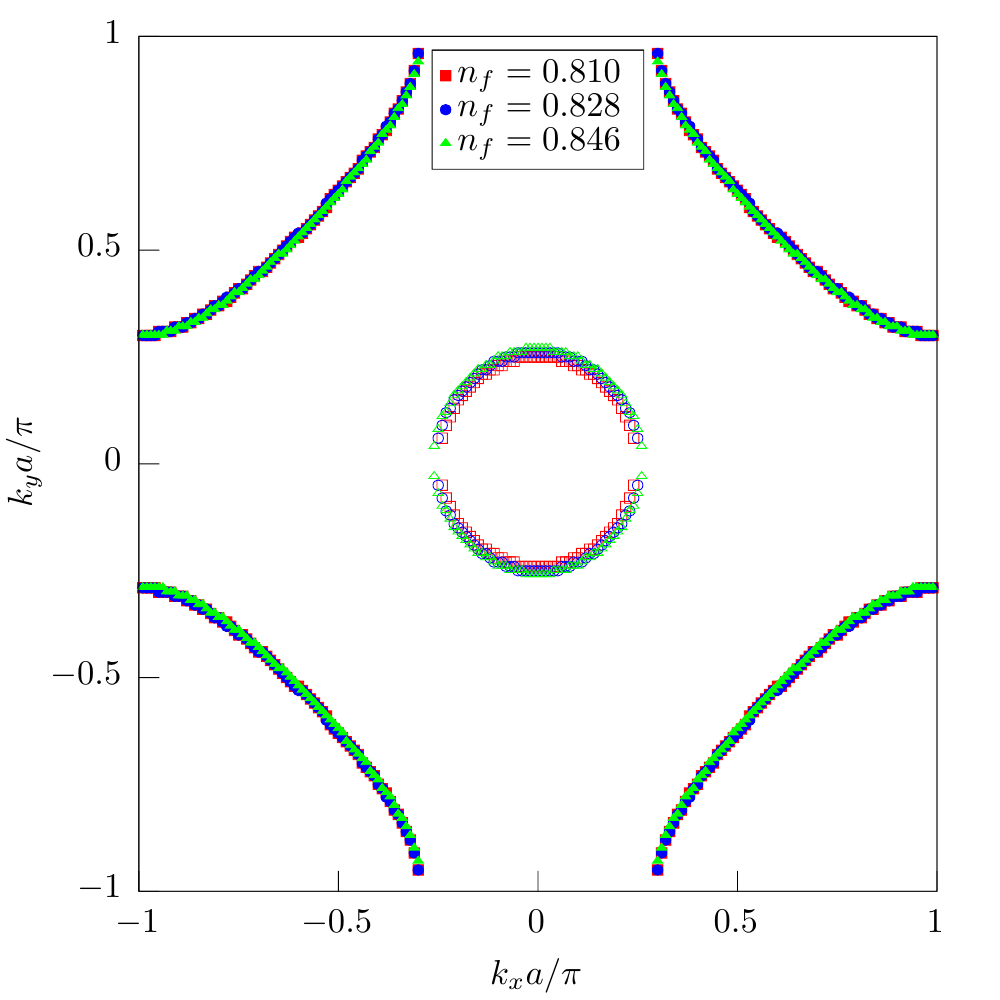}
\caption{FS's determined in the normal state at $T=0.001$ as functions of $n_{f}$ for fixed $t_{d}=0.1, J_{H}=0.0375, V=0.5$, and $n_{Tot}=1.473$. The FS's remain essentially unchanged within the range of $n_{f}$, although the pairing at zero-temperature undergoes a second-order transition from $s$- to $s+id$. }
\label{Fig:Appn_A_IV}
\end{figure}

In Fig.~\ref{Fig:Appn_A_IV} we show the evolution of the FS's determined in the normal state at $T=0.001$ as functions of $n_{f}$, for fixed $t_{d}=0.1, J_{H}=0.0375, n_{Tot}=1.473$, and $V=0.5$.  At zero-temperature there is a second-order phase transition from $s$-wave to $s+id$ pairing with increasing $n_{f}$. It is apparent that the FS does not significantly change, indicating that an additional change in the $f$ content of the bands is required for the emergence of this type of pairing.   
\enni

\begin{figure}[h!]
\centering
\includegraphics[width=1.0\columnwidth]{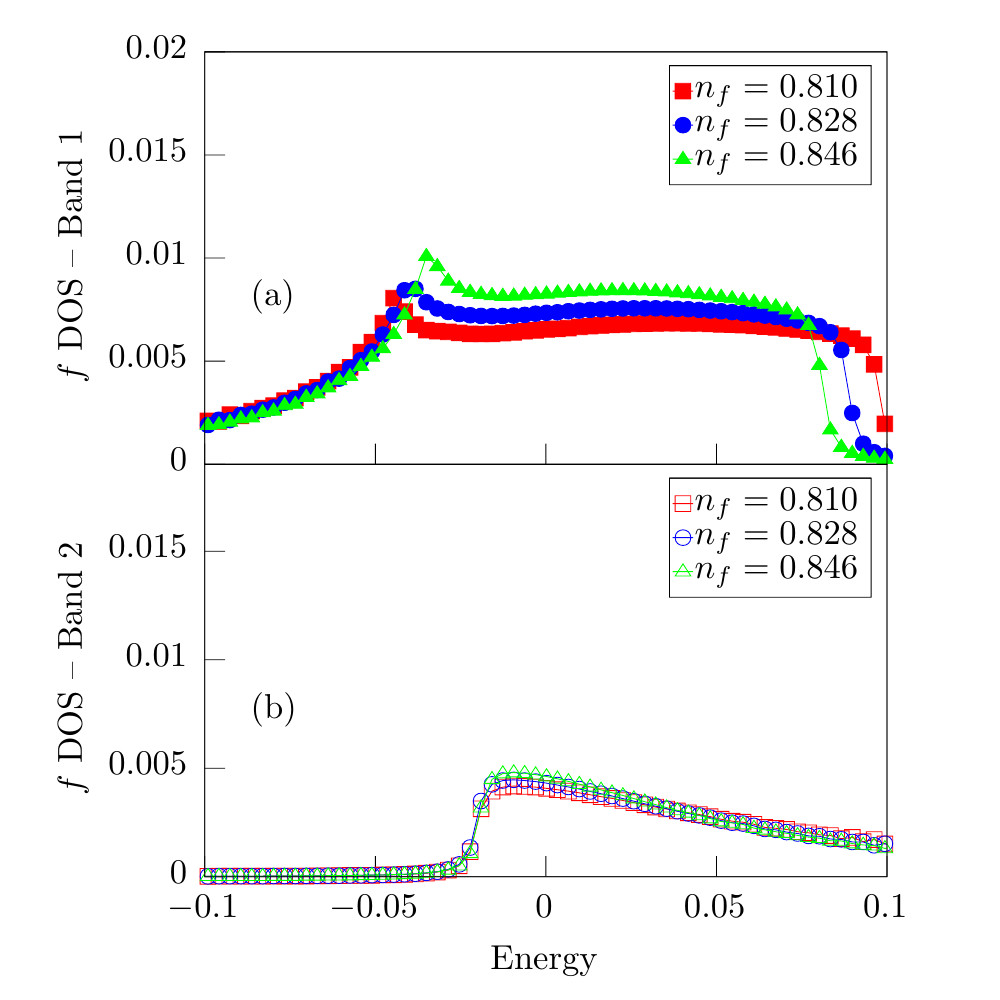}
\caption{$f$ DOS in the normal state at $T=0.001$ as functions of energy and $n_{f}$ for fixed $t_{d}=0.1, J_{H}=0.0375, n_{Tot}=1.473$, and $V=0.5$. A second-order transition between $s$-wave and $s+id$ occurs at zero-temperature across this range of $n_{f}$.}
\label{Fig:Appn_A_V}
\end{figure}

\enni The corresponding $f$ DOS is shown in Fig.~\ref{Fig:Appn_A_V}. There, the increase in the $f$ DOS near the Fermi level at zero-energy, together with a narrowing in the peaks, is apparent for increasing $n_{f}$.
\enni
\enni

\pagebreak


\begin{thebibliography}{38}%
\makeatletter
\providecommand \@ifxundefined [1]{%
 \@ifx{#1\undefined}
}%
\providecommand \@ifnum [1]{%
 \ifnum #1\expandafter \@firstoftwo
 \else \expandafter \@secondoftwo
 \fi
}%
\providecommand \@ifx [1]{%
 \ifx #1\expandafter \@firstoftwo
 \else \expandafter \@secondoftwo
 \fi
}%
\providecommand \natexlab [1]{#1}%
\providecommand \enquote  [1]{``#1''}%
\providecommand \bibnamefont  [1]{#1}%
\providecommand \bibfnamefont [1]{#1}%
\providecommand \citenamefont [1]{#1}%
\providecommand \href@noop [0]{\@secondoftwo}%
\providecommand \href [0]{\begingroup \@sanitize@url \@href}%
\providecommand \@href[1]{\@@startlink{#1}\@@href}%
\providecommand \@@href[1]{\endgroup#1\@@endlink}%
\providecommand \@sanitize@url [0]{\catcode `\\12\catcode `\$12\catcode
  `\&12\catcode `\#12\catcode `\^12\catcode `\_12\catcode `\%12\relax}%
\providecommand \@@startlink[1]{}%
\providecommand \@@endlink[0]{}%
\providecommand \url  [0]{\begingroup\@sanitize@url \@url }%
\providecommand \@url [1]{\endgroup\@href {#1}{\urlprefix }}%
\providecommand \urlprefix  [0]{URL }%
\providecommand \Eprint [0]{\href }%
\providecommand \doibase [0]{http://dx.doi.org/}%
\providecommand \selectlanguage [0]{\@gobble}%
\providecommand \bibinfo  [0]{\@secondoftwo}%
\providecommand \bibfield  [0]{\@secondoftwo}%
\providecommand \translation [1]{[#1]}%
\providecommand \BibitemOpen [0]{}%
\providecommand \bibitemStop [0]{}%
\providecommand \bibitemNoStop [0]{.\EOS\space}%
\providecommand \EOS [0]{\spacefactor3000\relax}%
\providecommand \BibitemShut  [1]{\csname bibitem#1\endcsname}%
\let\auto@bib@innerbib\@empty
\bibitem [{\citenamefont {Middey}\ \emph {et~al.}(2016)\citenamefont {Middey},
  \citenamefont {Chakhalian}, \citenamefont {Mahadevan}, \citenamefont
  {Freeland}, \citenamefont {Millis},\ and\ \citenamefont
  {Sarma}}]{Middey_ARMR2016}%
  \BibitemOpen
  \bibfield  {author} {\bibinfo {author} {\bibfnamefont {S.}~\bibnamefont
  {Middey}}, \bibinfo {author} {\bibfnamefont {J.}~\bibnamefont {Chakhalian}},
  \bibinfo {author} {\bibfnamefont {P.}~\bibnamefont {Mahadevan}}, \bibinfo
  {author} {\bibfnamefont {J.W.}\ \bibnamefont {Freeland}}, \bibinfo {author}
  {\bibfnamefont {A.J.}\ \bibnamefont {Millis}}, \ and\ \bibinfo {author}
  {\bibfnamefont {D.D.}\ \bibnamefont {Sarma}},\ }\bibfield  {title} {\enquote
  {\bibinfo {title} {Physics of ultrathin films and heterostructures of
  rare-earth nickelates},}\ }\href {\doibase
  10.1146/annurev-matsci-070115-032057} {\bibfield  {journal} {\bibinfo
  {journal} {Annual Review of Materials Research}\ }\textbf {\bibinfo {volume}
  {46}},\ \bibinfo {pages} {305} (\bibinfo {year} {2016})}\BibitemShut
  {NoStop}%
\bibitem [{\citenamefont {Li}\ \emph {et~al.}(2019)\citenamefont {Li},
  \citenamefont {Lee}, \citenamefont {Wang}, \citenamefont {Osada},
  \citenamefont {Crossley}, \citenamefont {Lee}, \citenamefont {Cui},\ and\
  \citenamefont {Hikita}}]{Hwang_Nature2019}%
  \BibitemOpen
  \bibfield  {author} {\bibinfo {author} {\bibfnamefont {Danfeng}\ \bibnamefont
  {Li}}, \bibinfo {author} {\bibfnamefont {Kyuho}\ \bibnamefont {Lee}},
  \bibinfo {author} {\bibfnamefont {Bai~Yang}\ \bibnamefont {Wang}}, \bibinfo
  {author} {\bibfnamefont {Motoki}\ \bibnamefont {Osada}}, \bibinfo {author}
  {\bibfnamefont {Samuel}\ \bibnamefont {Crossley}}, \bibinfo {author}
  {\bibfnamefont {Hye~Ryoung}\ \bibnamefont {Lee}}, \bibinfo {author}
  {\bibfnamefont {Yi}~\bibnamefont {Cui}}, \ and\ \bibinfo {author}
  {\bibfnamefont {Harold~Y.}\ \bibnamefont {Hikita}, \bibfnamefont
  {Yasuyuki~Hwang}},\ }\bibfield  {title} {\enquote {\bibinfo {title}
  {Superconductivity in an infinite-layer nickelate},}\ }\href {\doibase
  https://doi.org/10.1038/s41586-019-1496-5} {\bibfield  {journal} {\bibinfo
  {journal} {Nature}\ }\textbf {\bibinfo {volume} {572}},\ \bibinfo {pages}
  {624--627} (\bibinfo {year} {2019})}\BibitemShut {NoStop}%
\bibitem [{\citenamefont {Sawatzky}(2019)}]{Sawatzky_Nature2019}%
  \BibitemOpen
  \bibfield  {author} {\bibinfo {author} {\bibfnamefont {George~A.}\
  \bibnamefont {Sawatzky}},\ }\bibfield  {title} {\enquote {\bibinfo {title}
  {Superconductivity seen in a non-magnetic nickel oxide},}\ }\href {\doibase
  https://doi.org/10.1038/d41586-019-02518-3} {\bibfield  {journal} {\bibinfo
  {journal} {Nature}\ }\textbf {\bibinfo {volume} {572}},\ \bibinfo {pages}
  {592} (\bibinfo {year} {2019})}\BibitemShut {NoStop}%
\bibitem [{\citenamefont {{Zeng}}\ \emph {et~al.}()\citenamefont {{Zeng}},
  \citenamefont {{Sin Tang}}, \citenamefont {{Yin}}, \citenamefont {{Li}},
  \citenamefont {{Huang}}, \citenamefont {{Hu}}, \citenamefont {{Liu}},
  \citenamefont {{Omar}}, \citenamefont {{Jani}}, \citenamefont {{Shiuh Lim}},
  \citenamefont {{Han}}, \citenamefont {{Wan}}, \citenamefont {{Yang}},
  \citenamefont {{Wee}},\ and\ \citenamefont {{Ariando}}}]{Zheng_2020}%
  \BibitemOpen
  \bibfield  {author} {\bibinfo {author} {\bibfnamefont {Shengwei}\
  \bibnamefont {{Zeng}}}, \bibinfo {author} {\bibfnamefont {Chi}\ \bibnamefont
  {{Sin Tang}}}, \bibinfo {author} {\bibfnamefont {Xinmao}\ \bibnamefont
  {{Yin}}}, \bibinfo {author} {\bibfnamefont {Changjian}\ \bibnamefont {{Li}}},
  \bibinfo {author} {\bibfnamefont {Zhen}\ \bibnamefont {{Huang}}}, \bibinfo
  {author} {\bibfnamefont {Junxiong}\ \bibnamefont {{Hu}}}, \bibinfo {author}
  {\bibfnamefont {Wei}\ \bibnamefont {{Liu}}}, \bibinfo {author} {\bibfnamefont
  {Ganesh~Ji}\ \bibnamefont {{Omar}}}, \bibinfo {author} {\bibfnamefont
  {Hariom}\ \bibnamefont {{Jani}}}, \bibinfo {author} {\bibfnamefont {Zhi}\
  \bibnamefont {{Shiuh Lim}}}, \bibinfo {author} {\bibfnamefont {Kun}\
  \bibnamefont {{Han}}}, \bibinfo {author} {\bibfnamefont {Dongyang}\
  \bibnamefont {{Wan}}}, \bibinfo {author} {\bibfnamefont {Ping}\ \bibnamefont
  {{Yang}}}, \bibinfo {author} {\bibfnamefont {Andrew T.~S.}\ \bibnamefont
  {{Wee}}}, \ and\ \bibinfo {author} {\bibfnamefont {Ariando}\ \bibnamefont
  {{Ariando}}},\ }\bibfield  {title} {\enquote {\bibinfo {title} {{Phase
  diagram and superconducting dome of infinite-layer
  $\mathrm{Nd_{1-x}Sr_{x}NiO_{2}}$ thin films}},}\ }\href@noop {} {\ }\Eprint
  {http://arxiv.org/abs/2004.11281} {arXiv:2004.11281} \BibitemShut {NoStop}%
\bibitem [{\citenamefont {Hepting}\ \emph {et~al.}(2020)\citenamefont
  {Hepting}, \citenamefont {Li}, \citenamefont {Jia}, \citenamefont {Lu},
  \citenamefont {Paris}, \citenamefont {Tseng}, \citenamefont {Feng},
  \citenamefont {Osada}, \citenamefont {Been}, \citenamefont {Hikita},
  \citenamefont {Chuang}, \citenamefont {Hussain}, \citenamefont {Zhou},
  \citenamefont {Nag}, \citenamefont {Garcia-Fernandez}, \citenamefont {Rossi},
  \citenamefont {Huang}, \citenamefont {Huang}, \citenamefont {Shen},
  \citenamefont {Schmitt}, \citenamefont {Hwang}, \citenamefont {Moritz},
  \citenamefont {Zaanen}, \citenamefont {Devereaux},\ and\ \citenamefont
  {Lee}}]{deveraux}%
  \BibitemOpen
  \bibfield  {author} {\bibinfo {author} {\bibfnamefont {M.}~\bibnamefont
  {Hepting}}, \bibinfo {author} {\bibfnamefont {D.}~\bibnamefont {Li}},
  \bibinfo {author} {\bibfnamefont {C.~J.}\ \bibnamefont {Jia}}, \bibinfo
  {author} {\bibfnamefont {H.}~\bibnamefont {Lu}}, \bibinfo {author}
  {\bibfnamefont {E.}~\bibnamefont {Paris}}, \bibinfo {author} {\bibfnamefont
  {Y.}~\bibnamefont {Tseng}}, \bibinfo {author} {\bibfnamefont
  {X.}~\bibnamefont {Feng}}, \bibinfo {author} {\bibfnamefont {M.}~\bibnamefont
  {Osada}}, \bibinfo {author} {\bibfnamefont {E.}~\bibnamefont {Been}},
  \bibinfo {author} {\bibfnamefont {Y.}~\bibnamefont {Hikita}}, \bibinfo
  {author} {\bibfnamefont {Y.~D.}\ \bibnamefont {Chuang}}, \bibinfo {author}
  {\bibfnamefont {Z.}~\bibnamefont {Hussain}}, \bibinfo {author} {\bibfnamefont
  {K.~J.}\ \bibnamefont {Zhou}}, \bibinfo {author} {\bibfnamefont
  {A.}~\bibnamefont {Nag}}, \bibinfo {author} {\bibfnamefont {M.}~\bibnamefont
  {Garcia-Fernandez}}, \bibinfo {author} {\bibfnamefont {M.}~\bibnamefont
  {Rossi}}, \bibinfo {author} {\bibfnamefont {H.~Y.}\ \bibnamefont {Huang}},
  \bibinfo {author} {\bibfnamefont {D.~J.}\ \bibnamefont {Huang}}, \bibinfo
  {author} {\bibfnamefont {Z.~X.}\ \bibnamefont {Shen}}, \bibinfo {author}
  {\bibfnamefont {T.}~\bibnamefont {Schmitt}}, \bibinfo {author} {\bibfnamefont
  {H.~Y.}\ \bibnamefont {Hwang}}, \bibinfo {author} {\bibfnamefont
  {B.}~\bibnamefont {Moritz}}, \bibinfo {author} {\bibfnamefont
  {J.}~\bibnamefont {Zaanen}}, \bibinfo {author} {\bibfnamefont {T.~P.}\
  \bibnamefont {Devereaux}}, \ and\ \bibinfo {author} {\bibfnamefont {W.~S.}\
  \bibnamefont {Lee}},\ }\bibfield  {title} {\enquote {\bibinfo {title}
  {Electronic structure of the parent compound of superconducting
  infinite-layer nickelates},}\ }\href
  {https://doi.org/10.1038/s41563-019-0585-z} {\bibfield  {journal} {\bibinfo
  {journal} {Nature Materials}\ }\textbf {\bibinfo {volume} {19}},\ \bibinfo
  {pages} {381} (\bibinfo {year} {2020})}\BibitemShut {NoStop}%
\bibitem [{\citenamefont {Si}\ and\ \citenamefont
  {Steglich}(2010)}]{Si_Science2010}%
  \BibitemOpen
  \bibfield  {author} {\bibinfo {author} {\bibfnamefont {Qimiao}\ \bibnamefont
  {Si}}\ and\ \bibinfo {author} {\bibfnamefont {Frank}\ \bibnamefont
  {Steglich}},\ }\bibfield  {title} {\enquote {\bibinfo {title} {Heavy fermions
  and quantum phase transitions},}\ }\href {\doibase 10.1126/science.1191195}
  {\bibfield  {journal} {\bibinfo  {journal} {Science}\ }\textbf {\bibinfo
  {volume} {329}},\ \bibinfo {pages} {1161} (\bibinfo {year}
  {2010})}\BibitemShut {NoStop}%
\bibitem [{\citenamefont {Curro}\ \emph {et~al.}(2005)\citenamefont {Curro},
  \citenamefont {Caldwell}, \citenamefont {Bauer}, \citenamefont {Morales},
  \citenamefont {Graf}, \citenamefont {Bang}, \citenamefont {Balatsky},
  \citenamefont {D.},\ and\ \citenamefont {Sarrao}}]{Curro_Nature2005}%
  \BibitemOpen
  \bibfield  {author} {\bibinfo {author} {\bibfnamefont {N.~J.}\ \bibnamefont
  {Curro}}, \bibinfo {author} {\bibfnamefont {T.}~\bibnamefont {Caldwell}},
  \bibinfo {author} {\bibfnamefont {E.~D.}\ \bibnamefont {Bauer}}, \bibinfo
  {author} {\bibfnamefont {L.~A.}\ \bibnamefont {Morales}}, \bibinfo {author}
  {\bibfnamefont {M.~J.}\ \bibnamefont {Graf}}, \bibinfo {author}
  {\bibfnamefont {Y.}~\bibnamefont {Bang}}, \bibinfo {author} {\bibfnamefont
  {A.~V.}\ \bibnamefont {Balatsky}}, \bibinfo {author} {\bibfnamefont
  {Thompson~J.}\ \bibnamefont {D.}}, \ and\ \bibinfo {author} {\bibfnamefont
  {J.~L.}\ \bibnamefont {Sarrao}},\ }\bibfield  {title} {\enquote {\bibinfo
  {title} {{Unconventional superconductivity in PuCoGa$_5$}},}\ }\href
  {\doibase https://doi.org/10.1038/nature03428} {\bibfield  {journal}
  {\bibinfo  {journal} {Nature}\ }\textbf {\bibinfo {volume} {434}},\ \bibinfo
  {pages} {622} (\bibinfo {year} {2005})}\BibitemShut {NoStop}%
\bibitem [{\citenamefont {Yu}\ \emph {et~al.}(2014)\citenamefont {Yu},
  \citenamefont {Zhu},\ and\ \citenamefont {Si}}]{Yu_2014}%
  \BibitemOpen
  \bibfield  {author} {\bibinfo {author} {\bibfnamefont {Rong}\ \bibnamefont
  {Yu}}, \bibinfo {author} {\bibfnamefont {Jian-Xin}\ \bibnamefont {Zhu}}, \
  and\ \bibinfo {author} {\bibfnamefont {Qimiao}\ \bibnamefont {Si}},\
  }\bibfield  {title} {\enquote {\bibinfo {title} {{Orbital-selective
  superconductivity, gap anisotropy, and spin resonance excitations in a
  multiorbital $t$-${J}_{1}$-${J}_{2}$ model for iron pnictides}},}\
  }\href@noop {} {\bibfield  {journal} {\bibinfo  {journal} {Phys. Rev. B}\
  }\textbf {\bibinfo {volume} {89}},\ \bibinfo {pages} {024509} (\bibinfo
  {year} {2014})}\BibitemShut {NoStop}%
\bibitem [{\citenamefont {Yin}\ \emph {et~al.}(2014)\citenamefont {Yin},
  \citenamefont {Haule},\ and\ \citenamefont {Kotliar}}]{Yin_Haule_Kotliar}%
  \BibitemOpen
  \bibfield  {author} {\bibinfo {author} {\bibfnamefont {Z.~P.}\ \bibnamefont
  {Yin}}, \bibinfo {author} {\bibfnamefont {K.}~\bibnamefont {Haule}}, \ and\
  \bibinfo {author} {\bibfnamefont {G.}~\bibnamefont {Kotliar}},\ }\bibfield
  {title} {\enquote {\bibinfo {title} {Spin dynamics and orbital-antiphase
  pairing symmetry in iron-based superconductors},}\ }\href {\doibase
  10.1038/nphys3116} {\bibfield  {journal} {\bibinfo  {journal} {Nature
  Physics}\ }\textbf {\bibinfo {volume} {10}},\ \bibinfo {pages} {845}
  (\bibinfo {year} {2014})}\BibitemShut {NoStop}%
\bibitem [{\citenamefont {Ong}\ \emph {et~al.}(2016)\citenamefont {Ong},
  \citenamefont {Coleman},\ and\ \citenamefont {Schmalian}}]{Coleman}%
  \BibitemOpen
  \bibfield  {author} {\bibinfo {author} {\bibfnamefont {Tzen}\ \bibnamefont
  {Ong}}, \bibinfo {author} {\bibfnamefont {Piers}\ \bibnamefont {Coleman}}, \
  and\ \bibinfo {author} {\bibfnamefont {J{\"o}rg}\ \bibnamefont {Schmalian}},\
  }\bibfield  {title} {\enquote {\bibinfo {title} {Concealed d-wave pairs in
  the s{\textpm} condensate of iron-based superconductors},}\ }\href {\doibase
  10.1073/pnas.1523064113} {\bibfield  {journal} {\bibinfo  {journal} {Proc.
  Nat. Acad. Sci.}\ }\textbf {\bibinfo {volume} {113}},\ \bibinfo {pages}
  {5486} (\bibinfo {year} {2016})}\BibitemShut {NoStop}%
\bibitem [{\citenamefont {Nica}\ \emph {et~al.}(2017)\citenamefont {Nica},
  \citenamefont {Yu},\ and\ \citenamefont {Si}}]{Nica_Yu}%
  \BibitemOpen
  \bibfield  {author} {\bibinfo {author} {\bibfnamefont {Emilian~M.}\
  \bibnamefont {Nica}}, \bibinfo {author} {\bibfnamefont {Rong}\ \bibnamefont
  {Yu}}, \ and\ \bibinfo {author} {\bibfnamefont {Qimiao}\ \bibnamefont {Si}},\
  }\bibfield  {title} {\enquote {\bibinfo {title} {Orbital-selective pairing
  and superconductivity in iron selenides},}\ }\href@noop {} {\bibfield
  {journal} {\bibinfo  {journal} {npj Quantum Materials}\ }\textbf {\bibinfo
  {volume} {2}},\ \bibinfo {pages} {24} (\bibinfo {year} {2017})},\ \Eprint
  {http://arxiv.org/abs/arXiv:1505.04170} {arXiv:1505.04170} \BibitemShut
  {NoStop}%
\bibitem [{\citenamefont {Kreisel}\ \emph {et~al.}(2017)\citenamefont
  {Kreisel}, \citenamefont {Andersen}, \citenamefont {Sprau}, \citenamefont
  {Kostin}, \citenamefont {Davis},\ and\ \citenamefont
  {Hirschfeld}}]{Kreisel2017}%
  \BibitemOpen
  \bibfield  {author} {\bibinfo {author} {\bibfnamefont {Andreas}\ \bibnamefont
  {Kreisel}}, \bibinfo {author} {\bibfnamefont {Brian~M.}\ \bibnamefont
  {Andersen}}, \bibinfo {author} {\bibfnamefont {P.~O.}\ \bibnamefont {Sprau}},
  \bibinfo {author} {\bibfnamefont {A.}~\bibnamefont {Kostin}}, \bibinfo
  {author} {\bibfnamefont {J.~C.~S\'eamus}\ \bibnamefont {Davis}}, \ and\
  \bibinfo {author} {\bibfnamefont {P.~J.}\ \bibnamefont {Hirschfeld}},\
  }\bibfield  {title} {\enquote {\bibinfo {title} {Orbital selective pairing
  and gap structures of iron-based superconductors},}\ }\href {\doibase
  10.1103/PhysRevB.95.174504} {\bibfield  {journal} {\bibinfo  {journal} {Phys.
  Rev. B}\ }\textbf {\bibinfo {volume} {95}},\ \bibinfo {pages} {174504}
  (\bibinfo {year} {2017})}\BibitemShut {NoStop}%
\bibitem [{\citenamefont {Hu}\ \emph {et~al.}(2018)\citenamefont {Hu},
  \citenamefont {Yu}, \citenamefont {Nica}, \citenamefont {Zhu},\ and\
  \citenamefont {Si}}]{HY_Hu2019}%
  \BibitemOpen
  \bibfield  {author} {\bibinfo {author} {\bibfnamefont {Haoyu}\ \bibnamefont
  {Hu}}, \bibinfo {author} {\bibfnamefont {Rong}\ \bibnamefont {Yu}}, \bibinfo
  {author} {\bibfnamefont {Emilian~M.}\ \bibnamefont {Nica}}, \bibinfo {author}
  {\bibfnamefont {Jian-Xin}\ \bibnamefont {Zhu}}, \ and\ \bibinfo {author}
  {\bibfnamefont {Qimiao}\ \bibnamefont {Si}},\ }\bibfield  {title} {\enquote
  {\bibinfo {title} {Orbital-selective superconductivity in the nematic phase
  of {F}e{S}e},}\ }\href {\doibase 10.1103/PhysRevB.98.220503} {\bibfield
  {journal} {\bibinfo  {journal} {Phys. Rev. B}\ }\textbf {\bibinfo {volume}
  {98}},\ \bibinfo {pages} {220503} (\bibinfo {year} {2018})}\BibitemShut
  {NoStop}%
\bibitem [{\citenamefont {Zhang}\ \emph {et~al.}(2013)\citenamefont {Zhang},
  \citenamefont {Yu}, \citenamefont {Su}, \citenamefont {Song}, \citenamefont
  {Wang}, \citenamefont {Tan}, \citenamefont {Egami}, \citenamefont
  {Fernandez-Baca}, \citenamefont {Faulhaber}, \citenamefont {Si},\ and\
  \citenamefont {Dai}}]{C_Zhang2013}%
  \BibitemOpen
  \bibfield  {author} {\bibinfo {author} {\bibfnamefont {Chenglin}\
  \bibnamefont {Zhang}}, \bibinfo {author} {\bibfnamefont {Rong}\ \bibnamefont
  {Yu}}, \bibinfo {author} {\bibfnamefont {Yixi}\ \bibnamefont {Su}}, \bibinfo
  {author} {\bibfnamefont {Yu}~\bibnamefont {Song}}, \bibinfo {author}
  {\bibfnamefont {Miaoyin}\ \bibnamefont {Wang}}, \bibinfo {author}
  {\bibfnamefont {Guotai}\ \bibnamefont {Tan}}, \bibinfo {author}
  {\bibfnamefont {Takeshi}\ \bibnamefont {Egami}}, \bibinfo {author}
  {\bibfnamefont {J.~A.}\ \bibnamefont {Fernandez-Baca}}, \bibinfo {author}
  {\bibfnamefont {Enrico}\ \bibnamefont {Faulhaber}}, \bibinfo {author}
  {\bibfnamefont {Qimiao}\ \bibnamefont {Si}}, \ and\ \bibinfo {author}
  {\bibfnamefont {Pengcheng}\ \bibnamefont {Dai}},\ }\bibfield  {title}
  {\enquote {\bibinfo {title} {Measurement of a double neutron-spin resonance
  and an anisotropic energy gap for underdoped superconducting
  {N}a{F}e$_{0.985}${C}o$_{0.015}${A}s using inelastic neutron scattering},}\
  }\href {\doibase 10.1103/PhysRevLett.111.207002} {\bibfield  {journal}
  {\bibinfo  {journal} {Phys. Rev. Lett.}\ }\textbf {\bibinfo {volume} {111}},\
  \bibinfo {pages} {207002} (\bibinfo {year} {2013})}\BibitemShut {NoStop}%
\bibitem [{\citenamefont {Sprau}\ \emph {et~al.}(2017)\citenamefont {Sprau},
  \citenamefont {Kostin}, \citenamefont {Kreisel}, \citenamefont {B{\"o}hmer},
  \citenamefont {Taufour}, \citenamefont {Canfield}, \citenamefont {Mukherjee},
  \citenamefont {Hirschfeld}, \citenamefont {Andersen},\ and\ \citenamefont
  {Davis}}]{Sprau}%
  \BibitemOpen
  \bibfield  {author} {\bibinfo {author} {\bibfnamefont {P.~O.}\ \bibnamefont
  {Sprau}}, \bibinfo {author} {\bibfnamefont {A.}~\bibnamefont {Kostin}},
  \bibinfo {author} {\bibfnamefont {A.}~\bibnamefont {Kreisel}}, \bibinfo
  {author} {\bibfnamefont {A.~E.}\ \bibnamefont {B{\"o}hmer}}, \bibinfo
  {author} {\bibfnamefont {V.}~\bibnamefont {Taufour}}, \bibinfo {author}
  {\bibfnamefont {P.~C.}\ \bibnamefont {Canfield}}, \bibinfo {author}
  {\bibfnamefont {S.}~\bibnamefont {Mukherjee}}, \bibinfo {author}
  {\bibfnamefont {P.~J.}\ \bibnamefont {Hirschfeld}}, \bibinfo {author}
  {\bibfnamefont {B.~M.}\ \bibnamefont {Andersen}}, \ and\ \bibinfo {author}
  {\bibfnamefont {J.~C.~S{\'e}amus}\ \bibnamefont {Davis}},\ }\bibfield
  {title} {\enquote {\bibinfo {title} {Discovery of orbital-selective {C}ooper
  pairing in {F}e{S}e},}\ }\href {\doibase 10.1126/science.aal1575} {\bibfield
  {journal} {\bibinfo  {journal} {Science}\ }\textbf {\bibinfo {volume}
  {357}},\ \bibinfo {pages} {75} (\bibinfo {year} {2017})}\BibitemShut
  {NoStop}%
\bibitem [{\citenamefont {Pang}\ \emph {et~al.}(2018)\citenamefont {Pang},
  \citenamefont {Smidman}, \citenamefont {Zhang}, \citenamefont {Jiao},
  \citenamefont {Weng}, \citenamefont {Nica}, \citenamefont {Chen},
  \citenamefont {Jiang}, \citenamefont {Zhang}, \citenamefont {Xie},
  \citenamefont {Jeevan}, \citenamefont {Lee}, \citenamefont {Gegenwart},
  \citenamefont {Steglich}, \citenamefont {Si},\ and\ \citenamefont
  {Yuan}}]{Pang}%
  \BibitemOpen
  \bibfield  {author} {\bibinfo {author} {\bibfnamefont {Guiming}\ \bibnamefont
  {Pang}}, \bibinfo {author} {\bibfnamefont {Michael}\ \bibnamefont {Smidman}},
  \bibinfo {author} {\bibfnamefont {Jinglei}\ \bibnamefont {Zhang}}, \bibinfo
  {author} {\bibfnamefont {Lin}\ \bibnamefont {Jiao}}, \bibinfo {author}
  {\bibfnamefont {Zongfa}\ \bibnamefont {Weng}}, \bibinfo {author}
  {\bibfnamefont {Emilian~M.}\ \bibnamefont {Nica}}, \bibinfo {author}
  {\bibfnamefont {Ye}~\bibnamefont {Chen}}, \bibinfo {author} {\bibfnamefont
  {Wenbing}\ \bibnamefont {Jiang}}, \bibinfo {author} {\bibfnamefont {Yongjun}\
  \bibnamefont {Zhang}}, \bibinfo {author} {\bibfnamefont {Wu}~\bibnamefont
  {Xie}}, \bibinfo {author} {\bibfnamefont {Hirale~S.}\ \bibnamefont {Jeevan}},
  \bibinfo {author} {\bibfnamefont {Hanoh}\ \bibnamefont {Lee}}, \bibinfo
  {author} {\bibfnamefont {Philipp}\ \bibnamefont {Gegenwart}}, \bibinfo
  {author} {\bibfnamefont {Frank}\ \bibnamefont {Steglich}}, \bibinfo {author}
  {\bibfnamefont {Qimiao}\ \bibnamefont {Si}}, \ and\ \bibinfo {author}
  {\bibfnamefont {Huiqiu}\ \bibnamefont {Yuan}},\ }\bibfield  {title} {\enquote
  {\bibinfo {title} {Fully gapped d-wave superconductivity in
  {C}e{C}u$_2${S}i$_2$},}\ }\href {\doibase 10.1073/pnas.1720291115} {\bibfield
   {journal} {\bibinfo  {journal} {Proc. Nat. Acad. Sci.}\ }\textbf {\bibinfo
  {volume} {115}},\ \bibinfo {pages} {5343} (\bibinfo {year}
  {2018})}\BibitemShut {NoStop}%
\bibitem [{\citenamefont {Nica}\ and\ \citenamefont {Si}()}]{Nica_Si}%
  \BibitemOpen
  \bibfield  {author} {\bibinfo {author} {\bibfnamefont {Emilian~M.}\
  \bibnamefont {Nica}}\ and\ \bibinfo {author} {\bibfnamefont {Qimiao}\
  \bibnamefont {Si}},\ }\href@noop {} {\enquote {\bibinfo {title} {Multiorbital
  singlet pairing and d+d superconductivity},}\ }\Eprint
  {http://arxiv.org/abs/arXiv:1911.13274} {arXiv:1911.13274} \BibitemShut
  {NoStop}%
\bibitem [{\citenamefont {Smidman}\ \emph {et~al.}(2018)\citenamefont
  {Smidman}, \citenamefont {Stockert}, \citenamefont {Arndt}, \citenamefont
  {Pang}, \citenamefont {Jiao}, \citenamefont {Yuan}, \citenamefont {Vieyra},
  \citenamefont {Kitagawa}, \citenamefont {Ishida}, \citenamefont {Fujiwara},
  \citenamefont {Kobayashi}, \citenamefont {Schuberth}, \citenamefont
  {Tippmann}, \citenamefont {Steinke}, \citenamefont {Lausberg}, \citenamefont
  {Steppke}, \citenamefont {Brando}, \citenamefont {Pfau}, \citenamefont
  {Stockert}, \citenamefont {Sun}, \citenamefont {Friedemann}, \citenamefont
  {Wirth}, \citenamefont {Krellner}, \citenamefont {Kirchner}, \citenamefont
  {Nica}, \citenamefont {Yu}, \citenamefont {Si},\ and\ \citenamefont
  {Steglich}}]{Smidman}%
  \BibitemOpen
  \bibfield  {author} {\bibinfo {author} {\bibfnamefont {M.}~\bibnamefont
  {Smidman}}, \bibinfo {author} {\bibfnamefont {O.}~\bibnamefont {Stockert}},
  \bibinfo {author} {\bibfnamefont {J.}~\bibnamefont {Arndt}}, \bibinfo
  {author} {\bibfnamefont {G.~M.}\ \bibnamefont {Pang}}, \bibinfo {author}
  {\bibfnamefont {L.}~\bibnamefont {Jiao}}, \bibinfo {author} {\bibfnamefont
  {H.~Q.}\ \bibnamefont {Yuan}}, \bibinfo {author} {\bibfnamefont {H.~A.}\
  \bibnamefont {Vieyra}}, \bibinfo {author} {\bibfnamefont {S.}~\bibnamefont
  {Kitagawa}}, \bibinfo {author} {\bibfnamefont {K.}~\bibnamefont {Ishida}},
  \bibinfo {author} {\bibfnamefont {K.}~\bibnamefont {Fujiwara}}, \bibinfo
  {author} {\bibfnamefont {T.~C.}\ \bibnamefont {Kobayashi}}, \bibinfo {author}
  {\bibfnamefont {E.}~\bibnamefont {Schuberth}}, \bibinfo {author}
  {\bibfnamefont {M.}~\bibnamefont {Tippmann}}, \bibinfo {author}
  {\bibfnamefont {L.}~\bibnamefont {Steinke}}, \bibinfo {author} {\bibfnamefont
  {S.}~\bibnamefont {Lausberg}}, \bibinfo {author} {\bibfnamefont
  {A.}~\bibnamefont {Steppke}}, \bibinfo {author} {\bibfnamefont
  {M.}~\bibnamefont {Brando}}, \bibinfo {author} {\bibfnamefont
  {H.}~\bibnamefont {Pfau}}, \bibinfo {author} {\bibfnamefont {U.}~\bibnamefont
  {Stockert}}, \bibinfo {author} {\bibfnamefont {P.}~\bibnamefont {Sun}},
  \bibinfo {author} {\bibfnamefont {S.}~\bibnamefont {Friedemann}}, \bibinfo
  {author} {\bibfnamefont {S.}~\bibnamefont {Wirth}}, \bibinfo {author}
  {\bibfnamefont {C.}~\bibnamefont {Krellner}}, \bibinfo {author}
  {\bibfnamefont {S.}~\bibnamefont {Kirchner}}, \bibinfo {author}
  {\bibfnamefont {E.~M.}\ \bibnamefont {Nica}}, \bibinfo {author}
  {\bibfnamefont {R.}~\bibnamefont {Yu}}, \bibinfo {author} {\bibfnamefont
  {Q.}~\bibnamefont {Si}}, \ and\ \bibinfo {author} {\bibfnamefont
  {F.}~\bibnamefont {Steglich}},\ }\bibfield  {title} {\enquote {\bibinfo
  {title} {Interplay between unconventional superconductivity and heavy-fermion
  quantum criticality: {C}e{C}u$_{2}${S}i$_{2}$ versus
  {Y}b{R}h$_{2}${S}i$_{2}$},}\ }\href {\doibase 10.1080/14786435.2018.1511070}
  {\bibfield  {journal} {\bibinfo  {journal} {Philos. Mag.}\ }\textbf {\bibinfo
  {volume} {98}},\ \bibinfo {pages} {2930} (\bibinfo {year}
  {2018})}\BibitemShut {NoStop}%
\bibitem [{\citenamefont {Scheidt}\ \emph {et~al.}(1998)\citenamefont
  {Scheidt}, \citenamefont {Schreiner}, \citenamefont {Kumar},\ and\
  \citenamefont {Stewart}}]{Scheidt_PRB1998}%
  \BibitemOpen
  \bibfield  {author} {\bibinfo {author} {\bibfnamefont {E.-W.}\ \bibnamefont
  {Scheidt}}, \bibinfo {author} {\bibfnamefont {T.}~\bibnamefont {Schreiner}},
  \bibinfo {author} {\bibfnamefont {P.}~\bibnamefont {Kumar}}, \ and\ \bibinfo
  {author} {\bibfnamefont {G.~R.}\ \bibnamefont {Stewart}},\ }\bibfield
  {title} {\enquote {\bibinfo {title} {{Specific heat study in
  ${\mathrm{U}}_{1\ensuremath{-}x}{\mathrm{Th}}_{x}{\mathrm{Be}}_{13}:$
  Enormous $\ensuremath{\Delta}C$ and strong coupling at ${x=x}_{c1}$ and
  ${x}_{c2};$ Correlation between \ensuremath{\gamma} and unusual
  superconductivity}},}\ }\href {\doibase 10.1103/PhysRevB.58.15153} {\bibfield
   {journal} {\bibinfo  {journal} {Phys. Rev. B}\ }\textbf {\bibinfo {volume}
  {58}},\ \bibinfo {pages} {15153} (\bibinfo {year} {1998})}\BibitemShut
  {NoStop}%
\bibitem [{\citenamefont {White}\ \emph {et~al.}(2015)\citenamefont {White},
  \citenamefont {Thompson},\ and\ \citenamefont {Maple}}]{White_PhyicaC2015}%
  \BibitemOpen
  \bibfield  {author} {\bibinfo {author} {\bibfnamefont {B.D.}\ \bibnamefont
  {White}}, \bibinfo {author} {\bibfnamefont {J.D.}\ \bibnamefont {Thompson}},
  \ and\ \bibinfo {author} {\bibfnamefont {M.B.}\ \bibnamefont {Maple}},\
  }\bibfield  {title} {\enquote {\bibinfo {title} {Unconventional
  superconductivity in heavy-fermion compounds},}\ }\href
  {http://www.sciencedirect.com/science/article/pii/S0921453415000714}
  {\bibfield  {journal} {\bibinfo  {journal} {Physica C}\ }\textbf {\bibinfo
  {volume} {514}},\ \bibinfo {pages} {246} (\bibinfo {year} {2015})},\ \bibinfo
  {note} {superconducting Materials: Conventional, Unconventional and
  Undetermined}\BibitemShut {NoStop}%
\bibitem [{\citenamefont {Stewart}(2017)}]{Stewart_AdvPhys2017}%
  \BibitemOpen
  \bibfield  {author} {\bibinfo {author} {\bibfnamefont {G.~R.}\ \bibnamefont
  {Stewart}},\ }\bibfield  {title} {\enquote {\bibinfo {title} {Unconventional
  superconductivity},}\ }\href@noop {} {\bibfield  {journal} {\bibinfo
  {journal} {Adv. Phys.}\ }\textbf {\bibinfo {volume} {66}},\ \bibinfo {pages}
  {75} (\bibinfo {year} {2017})}\BibitemShut {NoStop}%
\bibitem [{\citenamefont {Stewart}(2019)}]{Stewart_2019}%
  \BibitemOpen
  \bibfield  {author} {\bibinfo {author} {\bibfnamefont {G.~R.}\ \bibnamefont
  {Stewart}},\ }\bibfield  {title} {\enquote {\bibinfo {title} {{UBe$_13$ and
  U$_{1-x}$Th$_x$Be$_{13}$: Unconventional Superconductors}},}\ }\href@noop {}
  {\bibfield  {journal} {\bibinfo  {journal} {J. Low. Temp. Phys.}\ }\textbf
  {\bibinfo {volume} {195}},\ \bibinfo {pages} {1} (\bibinfo {year}
  {2019})}\BibitemShut {NoStop}%
\bibitem [{\citenamefont {Kumar}\ and\ \citenamefont
  {Wolfle}(1987)}]{Kumar_PRL1987}%
  \BibitemOpen
  \bibfield  {author} {\bibinfo {author} {\bibfnamefont {P.}~\bibnamefont
  {Kumar}}\ and\ \bibinfo {author} {\bibfnamefont {P.}~\bibnamefont {Wolfle}},\
  }\bibfield  {title} {\enquote {\bibinfo {title} {{Two-component
  order-parameter model for pure and thorium-doped superconducting
  ${\mathrm{UBe}}_{13}$}},}\ }\href {\doibase 10.1103/PhysRevLett.59.1954}
  {\bibfield  {journal} {\bibinfo  {journal} {Phys. Rev. Lett.}\ }\textbf
  {\bibinfo {volume} {59}},\ \bibinfo {pages} {1954} (\bibinfo {year}
  {1987})}\BibitemShut {NoStop}%
\bibitem [{\citenamefont {Kotliar}(1988)}]{Kotliar_s_id}%
  \BibitemOpen
  \bibfield  {author} {\bibinfo {author} {\bibfnamefont {G.}~\bibnamefont
  {Kotliar}},\ }\bibfield  {title} {\enquote {\bibinfo {title} {{Resonating
  valence bonds and d-wave superconductivity}},}\ }\href@noop {} {\bibfield
  {journal} {\bibinfo  {journal} {Phys. Rev. B}\ }\textbf {\bibinfo {volume}
  {37}},\ \bibinfo {pages} {3664} (\bibinfo {year} {1988})}\BibitemShut
  {NoStop}%
\bibitem [{\citenamefont {Sachdev}\ and\ \citenamefont
  {Read}(1991)}]{Sachdev_Read}%
  \BibitemOpen
  \bibfield  {author} {\bibinfo {author} {\bibfnamefont {Subir}\ \bibnamefont
  {Sachdev}}\ and\ \bibinfo {author} {\bibfnamefont {N.}~\bibnamefont {Read}},\
  }\bibfield  {title} {\enquote {\bibinfo {title} {{Large N expansion for
  frustrated and doped quantum antiferromagnets}},}\ }\href@noop {} {\bibfield
  {journal} {\bibinfo  {journal} {Int. J. Mod. Phys. B}\ }\textbf {\bibinfo
  {volume} {5}},\ \bibinfo {pages} {219} (\bibinfo {year} {1991})}\BibitemShut
  {NoStop}%
\bibitem [{\citenamefont {Kotliar}\ and\ \citenamefont {Liu}(1988)}]{Kotliar}%
  \BibitemOpen
  \bibfield  {author} {\bibinfo {author} {\bibfnamefont {Gabriel}\ \bibnamefont
  {Kotliar}}\ and\ \bibinfo {author} {\bibfnamefont {Jialin}\ \bibnamefont
  {Liu}},\ }\bibfield  {title} {\enquote {\bibinfo {title} {{Superexchange
  mechanism and d-wave superconductivity}},}\ }\href {\doibase
  10.1103/PhysRevB.38.5142} {\bibfield  {journal} {\bibinfo  {journal} {Phys.
  Rev. B}\ }\textbf {\bibinfo {volume} {38}},\ \bibinfo {pages} {5142}
  (\bibinfo {year} {1988})}\BibitemShut {NoStop}%
\bibitem [{\citenamefont {S.}\ and\ \citenamefont
  {Lawrence}(2016)}]{Riseborough_2016}%
  \BibitemOpen
  \bibfield  {author} {\bibinfo {author} {\bibfnamefont {Riseborough~P.}\
  \bibnamefont {S.}}\ and\ \bibinfo {author} {\bibfnamefont {J.~M.}\
  \bibnamefont {Lawrence}},\ }\bibfield  {title} {\enquote {\bibinfo {title}
  {Mixed valent metals},}\ }\href@noop {} {\bibfield  {journal} {\bibinfo
  {journal} {Rep. Prog. Phys.}\ }\textbf {\bibinfo {volume} {79}},\ \bibinfo
  {pages} {084501} (\bibinfo {year} {2016})}\BibitemShut {NoStop}%
\bibitem [{\citenamefont {Coleman}\ \emph {et~al.}(1989)\citenamefont
  {Coleman}, ,\ and\ \citenamefont {Andrei}}]{Coleman_Andrei}%
  \BibitemOpen
  \bibfield  {author} {\bibinfo {author} {\bibfnamefont {P.}~\bibnamefont
  {Coleman}}, , \ and\ \bibinfo {author} {\bibfnamefont {N.}~\bibnamefont
  {Andrei}},\ }\bibfield  {title} {\enquote {\bibinfo {title}
  {{Kondo-stabilised spin liquids and heavy fermion superconductivity}},}\
  }\href@noop {} {\bibfield  {journal} {\bibinfo  {journal} {J.Phys.: Condens.
  Matter}\ }\textbf {\bibinfo {volume} {1}},\ \bibinfo {pages} {4057} (\bibinfo
  {year} {1989})}\BibitemShut {NoStop}%
\bibitem [{\citenamefont {Lee}\ \emph {et~al.}(2009)\citenamefont {Lee},
  \citenamefont {Zhang},\ and\ \citenamefont {Wu}}]{Lee_Zhang_Wu}%
  \BibitemOpen
  \bibfield  {author} {\bibinfo {author} {\bibfnamefont {Wei-Cheng}\
  \bibnamefont {Lee}}, \bibinfo {author} {\bibfnamefont {Shou-Cheng}\
  \bibnamefont {Zhang}}, \ and\ \bibinfo {author} {\bibfnamefont {Congjun}\
  \bibnamefont {Wu}},\ }\bibfield  {title} {\enquote {\bibinfo {title} {Pairing
  state with a time-reversal symmetry breaking in {F}e{A}s-based
  superconductors},}\ }\href {\doibase 10.1103/PhysRevLett.102.217002}
  {\bibfield  {journal} {\bibinfo  {journal} {Phys. Rev. Lett.}\ }\textbf
  {\bibinfo {volume} {102}},\ \bibinfo {pages} {217002} (\bibinfo {year}
  {2009})}\BibitemShut {NoStop}%
\bibitem [{\citenamefont {Andrei}\ and\ \citenamefont
  {Coleman}(1989)}]{Andrei_Coleman}%
  \BibitemOpen
  \bibfield  {author} {\bibinfo {author} {\bibfnamefont {N.}~\bibnamefont
  {Andrei}}\ and\ \bibinfo {author} {\bibfnamefont {P.}~\bibnamefont
  {Coleman}},\ }\bibfield  {title} {\enquote {\bibinfo {title} {{Cooper
  Instability in the Presence of a Spin Liquid}},}\ }\href@noop {} {\bibfield
  {journal} {\bibinfo  {journal} {Phys. Rev. Lett.}\ }\textbf {\bibinfo
  {volume} {62}},\ \bibinfo {pages} {595} (\bibinfo {year} {1989})}\BibitemShut
  {NoStop}%
\bibitem [{\citenamefont {Goswami}\ \emph {et~al.}(2010)\citenamefont
  {Goswami}, \citenamefont {Nikolic},\ and\ \citenamefont {Si}}]{Goswami}%
  \BibitemOpen
  \bibfield  {author} {\bibinfo {author} {\bibfnamefont {Pallab}\ \bibnamefont
  {Goswami}}, \bibinfo {author} {\bibfnamefont {Predrag}\ \bibnamefont
  {Nikolic}}, \ and\ \bibinfo {author} {\bibfnamefont {Qimiao}\ \bibnamefont
  {Si}},\ }\bibfield  {title} {\enquote {\bibinfo {title} {Superconductivity in
  multi-orbital $t-{J}_{1}-{J}_{2}$ model and its implications for iron
  pnictides},}\ }\href {\doibase 10.1209/0295-5075/91/37006} {\bibfield
  {journal} {\bibinfo  {journal} {Europhys. Lett.}\ }\textbf {\bibinfo {volume}
  {91}},\ \bibinfo {pages} {37006} (\bibinfo {year} {2010})}\BibitemShut
  {NoStop}%
\bibitem [{\citenamefont {Flint}\ \emph {et~al.}(2008)\citenamefont {Flint},
  \citenamefont {Dzero},\ and\ \citenamefont {Coleman}}]{Flint_Dzero_Coleman}%
  \BibitemOpen
  \bibfield  {author} {\bibinfo {author} {\bibfnamefont {Rebecca}\ \bibnamefont
  {Flint}}, \bibinfo {author} {\bibfnamefont {M.}~\bibnamefont {Dzero}}, \ and\
  \bibinfo {author} {\bibfnamefont {P.}~\bibnamefont {Coleman}},\ }\bibfield
  {title} {\enquote {\bibinfo {title} {{Heavy electrons and the symplectic
  symmetry of spin}},}\ }\href@noop {} {\bibfield  {journal} {\bibinfo
  {journal} {Nature Physics}\ }\textbf {\bibinfo {volume} {4}},\ \bibinfo
  {pages} {643} (\bibinfo {year} {2008})}\BibitemShut {NoStop}%
\bibitem [{\citenamefont {Vojta}\ \emph {et~al.}(2000)\citenamefont {Vojta},
  \citenamefont {Zhang},\ and\ \citenamefont {Sachdev}}]{Vojta_Sachdev}%
  \BibitemOpen
  \bibfield  {author} {\bibinfo {author} {\bibfnamefont {Matthias}\
  \bibnamefont {Vojta}}, \bibinfo {author} {\bibfnamefont {Ying}\ \bibnamefont
  {Zhang}}, \ and\ \bibinfo {author} {\bibfnamefont {Subir}\ \bibnamefont
  {Sachdev}},\ }\bibfield  {title} {\enquote {\bibinfo {title} {{Competing
  orders and quantum criticality in doped antiferromagnets}},}\ }\href@noop {}
  {\bibfield  {journal} {\bibinfo  {journal} {Phys. Rev. B}\ }\textbf {\bibinfo
  {volume} {62}},\ \bibinfo {pages} {6721} (\bibinfo {year}
  {2000})}\BibitemShut {NoStop}%
\bibitem [{\citenamefont {Coleman}(1987)}]{Coleman_1987}%
  \BibitemOpen
  \bibfield  {author} {\bibinfo {author} {\bibfnamefont {Piers}\ \bibnamefont
  {Coleman}},\ }\bibfield  {title} {\enquote {\bibinfo {title} {Mixed valence
  as an almost broken symmetry},}\ }\href {\doibase 10.1103/PhysRevB.35.5072}
  {\bibfield  {journal} {\bibinfo  {journal} {Phys. Rev. B}\ }\textbf {\bibinfo
  {volume} {35}},\ \bibinfo {pages} {5072} (\bibinfo {year}
  {1987})}\BibitemShut {NoStop}%
\bibitem [{\citenamefont {N.}\ and\ \citenamefont {Newns}(1983)}]{Read_Newns}%
  \BibitemOpen
  \bibfield  {author} {\bibinfo {author} {\bibfnamefont {Read}\ \bibnamefont
  {N.}}\ and\ \bibinfo {author} {\bibfnamefont {D.M.}\ \bibnamefont {Newns}},\
  }\bibfield  {title} {\enquote {\bibinfo {title} {{On the solution of the
  Coqblin-Schreiffer Hamiltonian by the large-N expansion technique}},}\
  }\href@noop {} {\bibfield  {journal} {\bibinfo  {journal} {J. Phys. C}\
  }\textbf {\bibinfo {volume} {16}},\ \bibinfo {pages} {3273} (\bibinfo {year}
  {1983})}\BibitemShut {NoStop}%
\bibitem [{\citenamefont {Lee}\ \emph {et~al.}(2006)\citenamefont {Lee},
  \citenamefont {Nagaosa},\ and\ \citenamefont {Wen}}]{PA_Lee_2006}%
  \BibitemOpen
  \bibfield  {author} {\bibinfo {author} {\bibfnamefont {Patrick~A.}\
  \bibnamefont {Lee}}, \bibinfo {author} {\bibfnamefont {Naoto}\ \bibnamefont
  {Nagaosa}}, \ and\ \bibinfo {author} {\bibfnamefont {Xiao-Gang}\ \bibnamefont
  {Wen}},\ }\bibfield  {title} {\enquote {\bibinfo {title} {{Doping a Mott
  insulator: Physics of high-temperature superconductivity}},}\ }\href@noop {}
  {\bibfield  {journal} {\bibinfo  {journal} {Rev. Mod. Phys.}\ }\textbf
  {\bibinfo {volume} {78}},\ \bibinfo {pages} {17} (\bibinfo {year}
  {2006})}\BibitemShut {NoStop}%
\bibitem [{\citenamefont {Wang}\ \emph {et~al.}()\citenamefont {Wang},
  \citenamefont {Zhang}, \citenamefont {Yang},\ and\ \citenamefont
  {Zhang}}]{Zhan_2020}%
  \BibitemOpen
  \bibfield  {author} {\bibinfo {author} {\bibfnamefont {Zhan}\ \bibnamefont
  {Wang}}, \bibinfo {author} {\bibfnamefont {Guang-Ming}\ \bibnamefont
  {Zhang}}, \bibinfo {author} {\bibfnamefont {Yi-feng}\ \bibnamefont {Yang}}, \
  and\ \bibinfo {author} {\bibfnamefont {Fu-Chun}\ \bibnamefont {Zhang}},\
  }\bibfield  {title} {\enquote {\bibinfo {title} {Distinct pairing symmetries
  of superconductivity in infinite-layer nickelates},}\ }\href@noop {} {\
  }\Eprint {http://arxiv.org/abs/arXiv:2006.15928} {arXiv:2006.15928}
  \BibitemShut {NoStop}%
\bibitem [{\citenamefont {Gu}\ \emph {et~al.}()\citenamefont {Gu},
  \citenamefont {Li}, \citenamefont {Wan}, \citenamefont {Li}, \citenamefont
  {Guo}, \citenamefont {Yang}, \citenamefont {Li}, \citenamefont {Zhu},
  \citenamefont {Pan}, \citenamefont {Nie},\ and\ \citenamefont
  {Wen}}]{Gu_2020}%
  \BibitemOpen
  \bibfield  {author} {\bibinfo {author} {\bibfnamefont {Qiangqiang}\
  \bibnamefont {Gu}}, \bibinfo {author} {\bibfnamefont {Yueying}\ \bibnamefont
  {Li}}, \bibinfo {author} {\bibfnamefont {Siyuan}\ \bibnamefont {Wan}},
  \bibinfo {author} {\bibfnamefont {Huazhou}\ \bibnamefont {Li}}, \bibinfo
  {author} {\bibfnamefont {Wei}\ \bibnamefont {Guo}}, \bibinfo {author}
  {\bibfnamefont {Huan}\ \bibnamefont {Yang}}, \bibinfo {author} {\bibfnamefont
  {Qing}\ \bibnamefont {Li}}, \bibinfo {author} {\bibfnamefont {Xiyu}\
  \bibnamefont {Zhu}}, \bibinfo {author} {\bibfnamefont {Xiaoqing}\
  \bibnamefont {Pan}}, \bibinfo {author} {\bibfnamefont {Yuefeng}\ \bibnamefont
  {Nie}}, \ and\ \bibinfo {author} {\bibfnamefont {Hai-Hu}\ \bibnamefont
  {Wen}},\ }\bibfield  {title} {\enquote {\bibinfo {title} {Two superconducting
  components with different symmetries in nd1-xsrxnio2 films},}\ }\href@noop {}
  {\ }\Eprint {http://arxiv.org/abs/arXiv:2006.13123} {arXiv:2006.13123}
  \BibitemShut {NoStop}%
\end{thebibliography}
%

\end{document}